\documentclass[conference]{IEEEtran}
\usepackage{algorithmic}
\usepackage{graphicx}
\usepackage{pifont}
\newcommand{\cmark}{\ding{51}}
\newcommand{\xmark}{\ding{55}}
\usepackage{textcomp}
\usepackage[dvipsnames]{xcolor}
\usepackage{xspace}
\usepackage{fontawesome}
\usepackage{multirow}
\usepackage{colortbl}
\usepackage{rotating}
\usepackage{tikz}
\usepackage{soul}
\usepackage{makecell}
\usepackage{url}
\usepackage{enumitem}
\usepackage{fancybox}
\usepackage{booktabs}
\usepackage{fontawesome}
\usepackage{setspace}               % for LINE SPACING
\usepackage{pifont}
\usepackage[many]{tcolorbox}    	% for COLORED BOXES (tikz and xcolor included)

\tcbset{
  summarybox/.style={
    enhanced,
    attach boxed title to top left={yshift=-0pt},
    boxed title style={
        colback=navy,
        arc=3pt 3pt 0pt 0pt,
        boxrule=0pt,
        colframe=navy,
        fontupper=\scriptsize\bfseries\color{white}
    },
    colback=blue!6!white,
    colframe=blue!30!black!40,
    arc=0pt 6pt 6pt 6pt,
    boxrule=0.4pt,
    fontupper=\small,
    left=6pt, right=6pt,
    top=4pt, bottom=4pt,
    before skip=4pt,
    after skip=4pt
    }
}
\usepackage[caption=false]{subfig}
\usepackage[export]{adjustbox}
\usepackage{amsmath,amsfonts}
\usepackage{MnSymbol}
\usepackage{color,colortbl,booktabs}
\usepackage[noadjust]{cite}
\usepackage{balance}
\usepackage{bm}
\usepackage{placeins}

\definecolor{navy}{RGB}{0, 32, 96}
\definecolor{fprow}{RGB}{219, 229, 241}

\definecolor{main}{HTML}{cccccc}    % setting main color to be used
\definecolor{sub}{HTML}{000000}     % setting sub color to be used

\definecolor{darkgreen}{rgb}{0.0, 0.5, 0.0}

\newcommand{\ie}{\emph{i.e.,}\xspace}
\newcommand{\eg}{\emph{e.g.,}\xspace}

\newcommand{\etal}{\emph{et~al.}\xspace}
\newcommand{\secref}[1]{Section~\ref{#1}\xspace}

\newcommand{\rev}[1]{\textcolor{black}{#1}}
\newboolean{showcomments}
\setboolean{showcomments}{true}

\ifthenelse{\boolean{showcomments}}
{\newcommand{\nb}[2]{
		\fbox{\bfseries\sffamily\scriptsize#1}
		{\sf\small$\blacktriangleright$\textit{#2}$\blacktriangleleft$}
	}
	
}
{\newcommand{\nb}[2]{}
	
}

\AtBeginDocument{%
	\providecommand\BibTeX{{%
			Bib\TeX}}}
\def\BibTeX{{\rm B\kern-.05em{\sc i\kern-.025em b}\kern-.08em
    T\kern-.1667em\lower.7ex\hbox{E}\kern-.125emX}}
\begin{document}
\raggedbottom

\title{Quantize with Confidence? An Empirical Study of Quantization for Code Generation}

%\author{\IEEEauthorblockN{Anonymous Author(s)}}

%  \author{%
%   \IEEEauthorblockN{Saima Afrin}
%   \IEEEauthorblockA{%
%     \textit{Department of Computer Science}\\
%     \textit{William \& Mary}\\
%     Williamsburg, VA, USA\\
%     safrin@wm.edu
%   }
%   \and
%   \IEEEauthorblockN{Md. Zahidul Haque}
%   \IEEEauthorblockA{%
%     \textit{Department of Computer Science}\\
%     \textit{William \& Mary}\\
%     Williamsburg, VA, USA\\
%     mhaque@wm.edu
%   }
%   \and
%   \IEEEauthorblockN{Antonio Mastropaolo}
%   \IEEEauthorblockA{%
%     \textit{Department of Computer Science}\\
%     \textit{William \& Mary}\\
%     Williamsburg, VA, USA\\
%     amastropaolo@wm.edu
%   }
% }

\author{%
  \IEEEauthorblockN{Saima Afrin}
  \IEEEauthorblockA{%
    \textit{AURA Lab}\\
    \textit{Department of Computer Science}\\
    \textit{William \& Mary}\\
    Williamsburg, VA, USA\\
    safrin@wm.edu
  }
  \and
  \IEEEauthorblockN{Md. Zahidul Haque}
  \IEEEauthorblockA{%
    \textit{AURA Lab}\\
    \textit{Department of Computer Science}\\
    \textit{William \& Mary}\\
    Williamsburg, VA, USA\\
    mhaque@wm.edu
  }
  \and
  \IEEEauthorblockN{Antonio Mastropaolo}
  \IEEEauthorblockA{%
    \textit{AURA Lab}\\
    \textit{Department of Computer Science}\\
    \textit{William \& Mary}\\
    Williamsburg, VA, USA\\
    amastropaolo@wm.edu
  }
}

\maketitle

% ============================================================
% Abstract
% ============================================================

\begin{abstract}
The growing adoption of local inference frameworks such as Ollama has made it increasingly common for developers to run large code models on laptops and other resource-constrained hardware. In these settings, post-training quantization is essential for reducing memory footprint and enabling practical deployment. However, practitioners currently lack clear guidance on how different quantization techniques affect the functional correctness and quality of generated code. In this paper, we empirically investigate the extent to which state-of-the-art quantization methods — including GPTQ, AWQ, QuIP\#, AQLM, BitsAndBytes, and GGUF, affect two large code model families representative of the current state of practice: Qwen2.5-Coder and CodeLlama. Using McEval and CoderEval, two comprehensive multilingual benchmarks spanning Python and Java, we evaluate two dimensions of the generated code: (i) functional correctness, measured via pass@1; and (ii) code quality aspects -- including maintainability, reliability, security, and structural complexity. We further introduce a novel analysis of quantization robustness under varying prompt complexity -- characterized by Shannon entropy and token length, a dimension that, to our knowledge, remains unexplored in prior work. From the achieved results, it emerges that quantization techniques differ meaningfully in their effect on performance and code quality, for instance AQLM consistently matches or exceeds the full-precision baseline, whereas QuIP\# exhibits the largest correctness degradation, particularly on complex prompts. Security attributes remain stable across model families, benchmarks, and programming languages, but sensitivity to prompt complexity varies across techniques. Overall, our findings provide practical guidance for selecting quantization strategies when deploying large code models on resource-constrained hardware. More broadly, they highlight the importance of evaluating quantized models beyond functional correctness to account for code quality and sensitivity to prompt complexity.
\end{abstract}


% !TEX root = ../main.tex
% % ============================================================
% Section I: Introduction
% ============================================================

\section{Introduction}
\label{sec:introduction}

In recent years, Software Engineering (SE) has experienced a significant transformation, largely driven by the integration of Artificial Intelligence (AI) techniques into development practices. Among these, Large Language Models (LLMs) have emerged as powerful tools for automating a wide range of SE-related tasks, enabling developers and practitioners to work more efficiently. Collectively referred to as Large Code Models (LCMs), these models have been successfully applied to various aspects of software development automation, including code generation~\cite{li2022competition, Mastropaolo:icse2023}, code summarization~\cite{ahmed2022fewshot, ahmed2024automatic}, code translation~\cite{fan2023large}, and test case generation~\cite{tufano2020unit}, showcasing their versatility and impact across the whole spectrum of activities characterizing the software development lifecycle.

However, the remarkable capabilities of LCMs come at a significant cost. Models with billions of parameters demand substantial computational resources for inference, and their deployment often requires expensive GPU infrastructure that is inaccessible to many practitioners -- particularly those in smaller organizations or academic settings who wish to deploy models locally for privacy, customization, or cost reasons~\cite{novelli2024generative, wu2024unveiling}. The environmental implications are equally concerning: training and operating these models at scale produces a considerable carbon footprint, raising questions about the long-term sustainability of current development trajectories~\cite{castano2023exploring, patterson2021carbon, strubell2020energy}.

In this context, \emph{quantization}~\cite{gholami2022survey}, the process of reducing the numerical precision of model parameters from 16- or 32-bit floating point to compact representations such as 4-bit integers -- has emerged as one of the most practical strategies for bridging the gap between model capability and deployment feasibility. Unlike other efficiency techniques such as knowledge distillation \cite{hinton2015distillingknowledge} or pruning \cite{daloisio2024compression}, quantization operates directly on pre-trained model weights without requiring retraining, making it especially attractive for practitioners who need to deploy existing models on resource-constrained hardware.

% The growing popularity of quantized models reflects this need in practice. Lightweight local inference frameworks such as Ollama\footnote{\url{https://github.com/ollama/ollama}} now enable developers to run large language models directly on personal machines, and recent analyses report hundreds of thousands of deployed Ollama instances worldwide, highlighting the rapid growth of local LLM deployments outside centralized cloud infrastructures. At the same time, model repositories such as Hugging Face host a rapidly expanding ecosystem of quantized checkpoints; for example, the widely used \emph{TheBloke} repository alone provides thousands of quantized model variants across multiple architectures and quantization formats\footnote{\url{https://huggingface.co/TheBloke}}.

The growing popularity of quantized models reflects this need in practice. Lightweight local inference frameworks such as Ollama~\cite{ollama_repo} now enable developers to run large language models directly on personal machines, and recent analyses report hundreds of thousands of deployed Ollama instances worldwide, highlighting the rapid growth of local LLM deployments outside centralized cloud infrastructures. At the same time, model repositories such as Hugging Face host a rapidly expanding ecosystem of quantized checkpoints; for example, the widely used \emph{TheBloke} repository alone provides thousands of quantized model variants across multiple architectures and quantization formats~\cite{thebloke_hf}.

While this rapidly growing ecosystem makes quantized models increasingly accessible, it also introduces substantial variability in the techniques used to compress them. As a result, an emerging body of research has begun to investigate how different quantization strategies affect the behavior of large code models and in which scenarios of code generation this behavior drifts compared to its full-precision counterpart. Early investigations into this question have yielded encouraging but incomplete answers. Wei~\etal~\cite{wei2023towards} demonstrated that 8-bit quantization can preserve code generation performance with improved energy efficiency. Giagnorio~\etal~\cite{giagnorio2025quantizing} pushed the frontier further, showing that 4-bit quantization reduces memory by approximately 70\% without significant degradation, establishing 4-bit as the practical operating point. Nyamsuren~\cite{nyamsuren2025evaluating} reinforced this finding across low-resource programming languages. More recently, Afrin~\etal~\cite{afrin2025quantization} took the first step beyond functional correctness by examining how quantization affects code quality attributes such as maintainability and structural complexity, finding that AWQ largely preserves these properties.
% \rev{Although informative, that study examined only AWQ and asked whether quantization preserves quality; in contrast, our work asks \emph{which} technique practitioners should choose, treating technique selection itself as the object of study and contrasting six representatives of a unified PTQ taxonomy.}

% \rev{While Afrin \etal\cite{afrin2025quantization} took a first step beyond functional correctness by examining differences between code generated by quantized and non-quantized models, all prior studies \cite{afrin2025quantization, giagnorio2025quantizing, wei2023greener} evaluate only a single quantization technique and report aggregated results across tasks. Though this does not represent an issue yet, more of a feature underpinning the design of the study. {\color{purple} However, quantization is, by construction, a lossy compression technique that reduces numerical precision and the model's representational capacity, both of which support reasoning over syntax, semantics, and task-specific patterns. \cite{}}. {\color{purple}This suggests that any loss in generated-code quality may be non-uniform, varying with both the quantization technique applied and the complexity of the input being processed.}
% {\color{purple}As a result, the existing experimental designs and aggregate metrics leave two questions unanswered: (i) \textbf{whether degradation differs across quantization techniques under comparable conditions}; and (ii) \textbf{whether any degradation is concentrated on harder inputs rather than averaged away across tasks.} Our work addresses both.}}

\rev{However, quantization itself is a fundamentally lossy compression: it reduces numerical precision and the model's representational capacity, both of which support reasoning over syntax, semantics, and task-specific patterns~\cite{dettmers2023spqr}. This suggests that any loss in generated-code quality may be non-uniform, varying with both the quantization technique applied and the complexity of the input being processed. Existing experimental designs, by averaging across techniques and tasks, may therefore obscure such systematic effects rather than reveal their absence.}

All prior work has established quantization as a viable compression strategy for code LLMs, though several important gaps remain. First, existing studies evaluate only a small subset of quantization techniques, despite the Post Training Quantization (PTQ) landscape now encompassing multiple algorithmic families with potentially different behaviors.

Second, most evaluations rely solely on pass@$k$, overlooking code quality aspects such as maintainability and structural complexity~\cite{siddiq2024quality, yetistiren2023evaluating}. Third, prior work treats benchmark tasks uniformly and does not examine whether quantization effects vary with prompt complexity, which we approximate using observable properties of the input prompt such as token length and information density (\ie Shannon Entropy \cite{shannon1948mathematical}), potentially placing different demands on compressed model representations.

To address these gaps, we conduct an empirical study of six widely used weight-only quantization techniques -- GPTQ, AWQ, QuIP\#, AQLM, BitsAndBytes, and GGUF -- applied at 4-bit precision to two large code model families. We evaluate their impact on functional correctness (pass@1) and code quality (maintainability, reliability, security, and structural complexity) across multiple benchmarks and programming languages, and further analyze how robustness varies with prompt complexity, approximated through token length and Shannon entropy.

Our results show that 4-bit quantization largely preserves functional correctness, but the choice of technique introduces meaningful variation. AQLM consistently matches or exceeds the full-precision baseline, whereas QuIP\# shows the largest degradation on complex tasks. Code quality effects are limited but observable: security remains stable, AWQ increases maintainability issues in Java, and BitsAndBytes produces the largest degradation in Python complexity metrics. Robustness to prompt complexity is also strongly model-dependent. These findings provide practical guidance for selecting quantization strategies and highlight the importance of evaluating quantized models beyond functional correctness. All artifacts are released to support replication~\cite{replication}.

The remainder of this paper is organized as follows. Section~\ref{sec:related} reviews background and related work. Section~\ref{sec:design} describes the study design. Section~\ref{sec:implementation} details the implementation and evaluation setup. Section~\ref{sec:results} presents the empirical findings. Section~\ref{sec:threats} discusses threats to validity, and Section~\ref{sec:conclusion} concludes the paper.

\section{Background and Related Work}
\label{sec:related}

\subsection{Efficiency in Large Code Models}
\label{sec:efficiency}

The deployment of billion-parameter large code models imposes substantial computational and energy demands, raising concerns about environmental sustainability~\cite{castano2023exploring, patterson2021carbon, strubell2020energy}, particularly for practitioners deploying models locally for privacy, customization, or cost reasons~\cite{novelli2024generative, wu2024unveiling}. Shi~\etal~\cite{shi2024efficient} identified four key areas for improving sustainability---data reduction, model-centric, system-centric, and program-centric approaches---while Shi~\etal~\cite{shi2024greening} demonstrated that model compression can reduce energy usage by up to 184$\times$ and carbon emissions by 157$\times$ with negligible effectiveness loss.

Several directions address these challenges. Parameter-Efficient Fine-Tuning (PEFT) updates only a small parameter subset via adapters~\cite{houlsby2019parameter}, LoRA~\cite{hu2022lora}, or prompt tuning~\cite{lester2021power, li2021prefix}, with promising results on code tasks~\cite{wang2023one, weyssow2023exploring, liu2023empirical, ayupov2022parameter}. Knowledge Distillation (KD) trains smaller models to emulate larger ones~\cite{hsieh2023distilling, chaudhary2023code, wei2023magicoder}, but still requires teacher model access. Pruning removes redundant parameters at the cost of potential performance degradation~\cite{daloisio2024compression}. Among these, quantization---a model-centric approach~\cite{shi2024efficient}---is particularly practical: it directly reduces the precision of pre-trained parameters without retraining, making it well-suited for deploying code LLMs on resource-constrained hardware.

\subsubsection{Quantization}
\label{sec:quantization-bg}

Quantization reduces memory footprint and computational cost by representing model parameters in lower-precision formats (\eg 4-bit integers) instead of standard 16/32-bit floating point~\cite{gholami2022survey, wang2024survey}. Quantization-Aware Training (QAT)~\cite{esser2020learned} integrates quantization into training but requires full retraining---prohibitively expensive for billion-parameter LLMs~\cite{liu2024llm, shen2024edgeqat}. Post-Training Quantization (PTQ)~\cite{cai2020zeroq} converts pre-trained models using only a small calibration dataset, making it the dominant approach for compressing large code models and the focus of our study.

We organize weight-only PTQ approaches into six categories (Table~\ref{tab:quant-techniques}). Four are algorithmic: \emph{second-order/Hessian-informed} methods (GPTQ~\cite{frantar2022gptq}, OWQ~\cite{lee2024owq}, SpQR~\cite{dettmers2024spqr}) minimize quantization error using approximate Hessian information; \emph{activation-aware saliency} methods (AWQ~\cite{lin2024awq}, SqueezeLLM~\cite{kim2024squeezellm}, OmniQuant~\cite{shao2024omniquant}) protect critical weight channels via activation-magnitude analysis; \emph{rotation/incoherence-based} methods (QuIP~\cite{chee2024quip}, QuIP\#~\cite{tseng2024quip}, QuaRot~\cite{ashkboos2024quarot}, FlatQuant~\cite{sun2024flatquant}) apply orthogonal transforms to spread error uniformly; and \emph{vector quantization} methods (AQLM~\cite{egiazarian2024extreme}, VPTQ~\cite{liu2024vptq}) encode weight groups through learned codebooks. Two are ecosystem-level: \emph{library/kernel-based} implementations (BitsAndBytes~\cite{dettmers2023qlora}, HQQ~\cite{badri2023hqq}) provide efficient low-bit kernels with seamless framework integration, and \emph{format/runtime-based} solutions offer standardized formats for CPU-friendly inference. Weight-and-activation quantization---compressing both weights and activations---is more challenging due to input-dependent outliers; methods such as ZeroQuant~\cite{yao2022zeroquant}, SmoothQuant~\cite{xiao2023smoothquant}, LLM.int8()~\cite{dettmers2022gpt3}, RPTQ~\cite{yuan2023rptq}, and PB-LLM~\cite{shang2024pbllm} remain less widely adopted for code tasks. Jin~\etal~\cite{jin2024comprehensive} corroborated that 4-bit quantization retains performance comparable to full-precision counterparts across instruction-tuned LLMs (7B--72B).

For our study, we select one approach from each category: GPTQ~\cite{frantar2022gptq} (second-order/Hessian-informed), AWQ~\cite{lin2024awq} (activation-aware saliency), QuIP\#~\cite{tseng2024quip} (rotation/incoherence-based), AQLM~\cite{egiazarian2024extreme} (vector quantization), BitsAndBytes~\cite{dettmers2023qlora} (library/kernel-based), and GGUF (format/runtime-based), ensuring broad coverage across four algorithmic strategies and two dominant deployment pathways, all widely supported in major inference frameworks and with publicly available implementations. Detailed descriptions and configurations are in Section~\ref{sec:design}.

\begin{table}[t]
\centering
\caption{Summary of post-training quantization approaches for LLMs. ``W'' denotes weight-only; ``W+A'' denotes weight-and-activation quantization. ``Code Tasks?'' indicates prior application to code-related tasks.}
\label{tab:quant-techniques}
\renewcommand{\arraystretch}{0.80}
\resizebox{0.6\columnwidth}{!}{%
\begin{tabular}{@{}lccc@{}}
\toprule
\textbf{Approach} & \textbf{Type} & \textbf{Bits} & \textbf{Code Tasks} \\
\midrule
\multicolumn{4}{@{}l}{\textit{Second-order / Hessian-informed}} \\
GPTQ~\cite{frantar2022gptq}           & W       & 2--8  & \xmark \\
OWQ~\cite{lee2024owq}                 & W       & 3, 4  & \xmark \\
SpQR~\cite{dettmers2024spqr}          & W       & 3--8  & \xmark \\
\midrule
\multicolumn{4}{@{}l}{\textit{Activation-aware saliency}} \\
AWQ~\cite{lin2024awq}                 & W       & 4, 8  & \cmark\rlap{\scriptsize~\cite{afrin2025quantization}} \\
SqueezeLLM~\cite{kim2024squeezellm}   & W       & 3, 4  & \xmark \\
OmniQuant~\cite{shao2024omniquant}    & W / W+A & 2--8  & \xmark \\
\midrule
\multicolumn{4}{@{}l}{\textit{Rotation / incoherence-based}} \\
QuIP~\cite{chee2024quip}              & W       & 2--4  & \xmark \\
QuIP\#~\cite{tseng2024quip}           & W       & 2--4  & \xmark \\
QuaRot~\cite{ashkboos2024quarot}      & W+A     & 4     & \xmark \\
FlatQuant~\cite{sun2024flatquant}     & W+A     & 4     & \xmark \\
\midrule
\multicolumn{4}{@{}l}{\textit{Vector quantization (codebook-based)}} \\
AQLM~\cite{egiazarian2024extreme}     & W       & 2--4  & \cmark\rlap{\scriptsize~\cite{giagnorio2025quantizing}} \\
VPTQ~\cite{liu2024vptq}               & W       & 2--4  & \xmark \\
\midrule
\multicolumn{4}{@{}l}{\textit{Library / kernel-based}} \\
HQQ~\cite{badri2023hqq}               & W       & 2--8  & \xmark \\
BitsAndBytes~\cite{dettmers2023qlora} & W       & 4, 8  & \xmark \\
\midrule
\multicolumn{4}{@{}l}{\textit{Format / runtime-based}} \\
GGUF                                  & W       & 2--8  & \cmark\rlap{\scriptsize~\cite{nyamsuren2025evaluating}} \\
\midrule
\multicolumn{4}{@{}l}{\textit{Weight-and-activation}} \\
ZeroQuant~\cite{yao2022zeroquant}     & W+A     & 4, 8  & \xmark \\
SmoothQuant~\cite{xiao2023smoothquant}& W+A     & 8     & \xmark \\
LLM.int8()~\cite{dettmers2022gpt3}   & W+A     & 8     & \xmark \\
RPTQ~\cite{yuan2023rptq}              & W+A     & 4, 8  & \xmark \\
PB-LLM~\cite{shang2024pbllm}          & W+A     & 1, 2  & \xmark \\
\bottomrule
\end{tabular}%
}
\end{table}

\subsubsection{Quantization for Code Models and Code Quality}
\label{sec:quant-code}

While quantization has been extensively studied for general NLP tasks, its application to code-related activities remains comparatively nascent. Wei~\etal~\cite{wei2023towards} conducted the first large-scale investigation, examining 8-bit quantization on PLBART~\cite{ahmad2021unified}, CodeT5~\cite{wang2021codet5}, InCoder~\cite{fried2023incoder}, and CodeGen~\cite{nijkamp2023codegen} for code generation and summarization, finding that quantized models achieve improved energy efficiency with only marginal performance loss. Giagnorio~\etal~\cite{giagnorio2025quantizing} extended this to larger models (CodeLlama up to 34B, DeepSeek-Coder up to 33B) using AQLM at 2-bit precision, showing that 4-bit quantization reduces memory by approximately 70\% without significant degradation, while more extreme levels (2--3 bits) incur notable drops that can be partially mitigated through code-specific calibration and post-quantization fine-tuning. Nyamsuren~\cite{nyamsuren2025evaluating} evaluated five 7B models quantized in GGUF format for Lua code generation on consumer hardware, reinforcing the 4-bit trade-off while showing that degradation is more pronounced for low-resource languages. In another work~\cite{afrin2025quantization} presented the first study to go beyond functional correctness and examine quantization's impact on code \emph{quality}, finding that AWQ at 4-bit largely preserves quality metrics captured by static analysis tools.

Beyond quantization, a growing body of research recognizes that functional correctness alone is insufficient to characterize the practical utility of generated code~\cite{siddiq2023generate, yetistiren2023evaluating}. Siddiq and Santos~\cite{siddiq2024quality} found that LLM-generated code frequently contains quality issues despite being syntactically valid, while Liu~\etal~\cite{liu2024refining} and Kharma~\etal~\cite{kharma2025security} examined iterative refinement and security vulnerabilities, respectively. Yeti\c{s}tiren~\etal~\cite{yetistiren2023evaluating} evaluated code quality across multiple dimensions including maintainability and reliability.

Despite this progress, existing studies have predominantly focused on a single quantization technique per study, with pass@$k$ as the primary evaluation criterion. Afrin~\etal~\cite{afrin2025quantization} took the first step toward examining quantization's impact on code quality but were limited to AWQ alone. The present study addresses this gap by comparing six quantization techniques and assessing their differential effects on both functional correctness and code quality.
% !TEX root = ../main.tex
% % ============================================================
% Section 3: Study Methodology
% ============================================================

\section{Study Methodology}
\label{sec:design}

%This study conducts a comprehensive empirical investigation into how different quantization techniques affect the functional correctness, quality, and robustness of code generated by large code models. With a growing number of quantization methods available to practitioners---each built on distinct algorithmic foundations and compression strategies---it is essential to understand whether the choice of technique meaningfully influences the resulting code. To explore this, we evaluate six widely adopted weight-only quantization approaches applied at 4-bit precision and examine their impact across multiple dimensions of code generation.

We evaluate six widely adopted weight-only quantization approaches---GPTQ, AWQ, QuIP\#, AQLM, BitsAndBytes, and GGUF---applied at 4-bit precision to Qwen2.5-Coder-7B and CodeLlama-7B, and structure our investigation around the following research questions:

\begin{itemize}

    \item \textbf{RQ$_1$:} \emph{How do different quantization techniques impact the functional correctness of code generated by large code models?}

    In this RQ, we assess functional correctness using the pass@1 metric across two benchmarks (McEval \cite{chai2024mceval} and CoderEval \cite{yu2024codereval}) and two programming languages (Python and Java). By contrasting each quantized variant against its full-precision baseline, we aim to determine whether certain quantization approaches preserve correctness more effectively than others, or whether code generation capabilities degrade uniformly across compression strategies.

    \item \textbf{RQ$_2$:} \emph{How do different quantization techniques impact the quality attributes of automatically generated code?}

    Beyond correctness, RQ$_2$ examines maintainability, reliability, security, and structural complexity using SonarCloud \cite{sonarcloud} static analysis. Code that passes test cases may still exhibit excessive complexity or subtle defects that diminish its practical value~\cite{siddiq2024quality}; we compare quality attributes across all six approaches to identify whether certain methods introduce more pronounced degradation.

    \item \textbf{RQ$_3$:} \emph{Does input prompt complexity influence quantization-induced correctness degradation?}

    We hypothesize that more complex prompts---longer inputs with higher information density---may be disproportionately affected by weight compression. To test this, we evaluate all six approaches on three benchmarks forming a natural complexity gradient: McEval \cite{chai2024mceval} (short, simple prompts), CoderEval \cite{yu2024codereval} (mid-length, context-dependent), and BigCodeBench \cite{zhuo2024bigcodebench} (long, multi-library orchestration). This analysis is conducted on Python only. We characterize complexity using token length, and Shannon entropy \cite{shannon1948mathematical}, and examine whether correctness degradation correlates with these measures.

\end{itemize}

The remainder of this section describes our experimental setup, covering the selection of code generation models (\secref{sub:models}), the evaluation benchmarks (\secref{sub:benchmarks}), the quantization approaches under investigation (\secref{sub:quantization}), and the static analysis tools employed for quality assessment (\secref{sub:tools}).

\smallskip

% ----------------------------------------------------------
\subsection{Code Models and Dataset}
\label{sec:model-data}

This section outlines the code model families used in our experiments and the benchmarks employed to evaluate code generation across multiple programming languages.

\subsubsection{Code Models}
\label{sub:models}

We employ two well-established code model families: Qwen2.5-Coder~\cite{hui2024qwen2} and CodeLlama~\cite{roziere2023code}. Both have been extensively adopted in code generation and quantization research~\cite{li2023structured, coignion2024performance, ren2024reflectioncoder, afrin2025resource, afrin2025quantization, giagnorio2025quantizing}. Following established practices~\cite{afrin2025resource, li2023instructcoder, giagnorio2025quantizing, afrin2025quantization}, we use the instruction-tuned variant of each model.

\smallskip
\noindent\textbf{Qwen2.5-Coder~\cite{hui2024qwen2}}
  is a code-specialized adaptation of the Qwen2.5 architecture, pre-trained on over 5.5 trillion tokens comprising 70\% code, 20\% natural language, and 10\% mathematical content~\cite{hui2024qwen2}. 
  The family spans 0.5B to 32B parameters and has demonstrated strong performance across diverse code tasks~\cite{zhuo2024bigcodebench, quan2025codeelo}.
  %The family spans 0.5B to 32B parameters\footnote{\url{https://huggingface.co/collections/Qwen/qwen25-coder}} and has demonstrated strong performance across diverse code tasks~\cite{zhuo2024bigcodebench, quan2025codeelo}.

\smallskip
%\noindent\textbf{CodeLlama~\cite{roziere2023code}}
%is built on Llama-2~\cite{touvron2023llama} and further trained on 500 billion tokens of natural language and source code. It offers general-purpose and instruction-tuned variants\footnote{\url{https://huggingface.co/codellama}} ranging from 7B to 70B parameters.

\noindent\textbf{CodeLlama~\cite{roziere2023code}}
is built on Llama-2~\cite{touvron2023llama} and further trained on 500 billion tokens of natural language and source code. It offers general-purpose and instruction-tuned variants ranging from 7B to 70B parameters.

\smallskip
For both families, we restrict experiments to the \textbf{7B variant}. Prior research shows that correctness and quality improve with model size but with diminishing returns~\cite{afrin2025quantization, wei2023towards, giagnorio2025quantizing}; the 7B configuration is representative of the resource-constrained settings where quantization is most commonly applied. Constraining to a single size also keeps the combinatorial experimental space (six approaches $\times$ two benchmarks $\times$ two languages, plus a third benchmark for RQ$_3$) manageable while enabling deeper cross-technique analysis.

\smallskip
\subsubsection{Evaluation Benchmarks}
\label{sub:benchmarks}
We employ three code generation benchmarks.
\rev{McEval~\cite{chai2024mceval} and CoderEval~\cite{yu2024codereval} are used in both Python and Java for RQ$_1$ and RQ$_2$, and in Python for RQ$_3$. BigCodeBench~\cite{zhuo2024bigcodebench} is used in Python only and contributes to all three RQs, additionally spanning a gradient of input complexity (Python only) in RQ$_3$}.
%McEval~\cite{chai2024mceval} and CoderEval~\cite{yu2024codereval} serve as the primary benchmarks for RQ$_1$ and RQ$_2$ (Python and Java), while all three---including BigCodeBench~\cite{zhuo2024bigcodebench}---are used in RQ$_3$ to span a gradient of input complexity (Python only).

\smallskip
\noindent\textbf{McEval~\cite{chai2024mceval}} provides human-annotated, multilingual coding tasks with function signatures, docstrings, and test cases at varying difficulty levels. We use the Python (42 tasks) and Java (53 tasks) subsets. McEval represents the simplest end of our complexity spectrum, with concise function-level prompts.

\smallskip
\noindent\textbf{CoderEval~\cite{yu2024codereval}} comprises 230 Python and 230 Java tasks from real-world open-source projects, spanning six levels of context dependency---from self-contained functions to project-level contexts. Unlike standalone-only benchmarks, it reflects pragmatic generation scenarios and represents mid-length input complexity with richer contextual information.
\rev{We use the filtered CoderEval subset of Crupi~\etal~\cite{crupi2025effectiveness} (190 Python and 184 Java tasks), which removes tasks with unreliable test suites.}
\rev{While CoderEval can include file- or project-level context, the natural-language descriptions themselves are typically concise---the contextual code is supplied separately---explaining the relatively low mean prompt length in Table~\ref{tab:entropy-buckets}.}

\smallskip

% ----------------------------------------------------------
\subsection{Selected Quantization Approaches}
\label{sub:quantization}

As described in Section~\ref{sec:quantization-bg}, we select six weight-only PTQ approaches---one from each category in our taxonomy: \textbf{GPTQ}~\cite{frantar2022gptq} (second-order/Hessian-informed), \textbf{AWQ}~\cite{lin2024awq} (activation-aware saliency), \textbf{QuIP\#}~\cite{tseng2024quip} (rotation/incoherence-based), \textbf{AQLM}~\cite{egiazarian2024extreme} (vector quantization), \textbf{BitsAndBytes}~\cite{dettmers2023qlora} (library/kernel-based), and \textbf{GGUF} (format/runtime-based). All models are quantized to 4-bit precision, the most widely adopted operating point for practical LLM deployment~\cite{jin2024comprehensive}. Below, we provide a brief description of each approach and its configuration in our experiments.

\smallskip
\noindent\textbf{GPTQ}~\cite{frantar2022gptq} performs layer-wise weight quantization using approximate second-order (Hessian) information. After each column of the weight matrix is quantized, the remaining weights are adjusted to minimize the overall output reconstruction error. We use 4-bit quantization with a group size of 128.

\smallskip
\noindent\textbf{AWQ}~\cite{lin2024awq} identifies salient weight channels by analyzing activation magnitudes and applies per-channel scaling to protect them before quantization. This activation-aware approach preserves the weight channels that contribute most to model accuracy. We apply 4-bit quantization with the default AWQ configuration.

\smallskip
\noindent\textbf{QuIP\#}~\cite{tseng2024quip} uses randomized Hadamard transforms to make the weight matrices incoherent, spreading quantization error uniformly across dimensions, and employs E8 lattice codebooks for efficient encoding. 
%We apply 4-bit quantization using the E8P12RVQ4B codebook.

\smallskip
\noindent\textbf{AQLM}~\cite{egiazarian2024extreme} employs multi-codebook additive quantization, where groups of weights are jointly encoded through learned codebooks optimized across entire layer blocks. We use the 4-bit configuration with two codebooks.

\smallskip
\noindent\textbf{BitsAndBytes}~\cite{dettmers2023qlora} provides efficient low-bit quantization kernels with seamless integration into the Hugging Face ecosystem. We use 4-bit NormalFloat (NF4) quantization with double quantization enabled.

\smallskip
%\noindent\textbf{GGUF}\footnote{\url{https://github.com/ggerganov/ggml/blob/master/docs/gguf.md}} is a standardized format for distributing quantized model weights, enabling CPU-friendly inference via the llama.cpp ecosystem. We use the Q4\_K\_M quantization variant, which applies mixed-precision 4-bit quantization with medium-sized k-quant blocks.
\noindent\textbf{GGUF}~\cite{gguf_spec} is a standardized format for distributing quantized model weights, enabling CPU-friendly inference via the llama.cpp ecosystem. We use the Q4\_K\_M quantization variant, which applies mixed-precision 4-bit quantization with medium-sized k-quant blocks.

\smallskip
For consistency, we utilize pre-quantized model checkpoints available on Hugging Face where possible, and quantize models locally using official implementations when pre-quantized versions are unavailable.

\smallskip

\subsection{Static Analysis--Based Evaluation of Code Quality and Quantization Robustness}
\label{sub:tools}

Assessing the quality of generated code spans multiple dimensions---syntax validity, coding style, code smells, reliability, maintainability, and security~\cite{siddiq2023generate, yetistiren2023evaluating}. While various static analysis tools target these attributes, not all are necessary when their outputs converge on consistent findings.

For this study, we adopt \textbf{SonarCloud}~\cite{sonarcloud} as the sole static analysis tool for both Python and Java. SonarCloud identifies bugs, code smells, security vulnerabilities, and code duplications within a single, language-agnostic platform~\cite{kharma2025security}---well suited to our cross-language evaluation.

Rather than supplementing SonarCloud with language-specific tools---such as Pylint \cite{pylint} and Flake8 \cite{flake8} for Python or PMD \cite{pmd} for Java---we adopt a consolidated strategy. This is grounded in empirical evidence from Afrin~\etal~\cite{afrin2025quantization}, who found strong consistency between quality trends reported by SonarCloud and those captured by Pylint, Flake8, and PMD, with no meaningful divergence in conclusions. Given this alignment, additional tools would contribute redundancy without different insights.

To better understand how robustly quantized models maintain their code generation capabilities as task demands increase, we complement the quality assessment with an analysis of \emph{quantization robustness} across input complexity levels. We characterize each benchmark's prompts using token length, and Shannon entropy \cite{shannon1948mathematical}\rev{---two complementary, content-agnostic measures motivated by prior code-LLM evaluation work~\cite{liu2024fail} and by NLP literature on entropy as a measure of textual information density~\cite{genzel2002entropy}}---and examine whether quantization-induced correctness degradation, measured as the pass@1 difference between full-precision and quantized models---correlates with these complexity measures.

\section{Implementation and Evaluation}
\label{sec:implementation}

\subsection{Experimental Setup and Environment}
\label{sec:setup}

For each model family (Qwen2.5-Coder-7B-Instruct and CodeLlama-7B-Instruct), we prepare seven configurations: one full-precision (FP16) baseline and six 4-bit quantized variants (GPTQ, AWQ, QuIP\#, AQLM, BitsAndBytes, GGUF). 
%Each configuration generates solutions for all benchmark tasks---McEval and CoderEval in Python and Java (RQ$_1$, RQ$_2$), with BigCodeBench (Python) added for RQ$_3$. 
Each configuration generates solutions for all benchmark tasks \rev{across the three RQs---McEval and CoderEval are used in both Python and Java for RQ$_1$ and RQ$_2$ and in Python for RQ$_3$; BigCodeBench is used in Python only and contributes to all three RQs}.
Generated code is evaluated for functional correctness (pass@1~\cite{chen:arxiv2021}) within a sandboxed Docker environment and for code quality via SonarCloud metrics. We set temperature to 0 for deterministic outputs and cap input tokens at 1024, consistent with related work~\cite{wang2023review, fakhoury2024llm}. All experiments ran on Ubuntu 22.04.5 LTS with four NVIDIA L40S GPUs (48GB each).

\subsection{Model Quantization}
\label{sec:quantization-impl}

For AWQ, GPTQ, GGUF, and BitsAndBytes, we used pre-quantized checkpoints from Hugging Face. For AQLM and QuIP\#, we performed quantization using official pipelines with WikiText-2~\cite{merity2016pointer} as the calibration corpus, a standard choice for PTQ~\cite{frantar2022gptq, lin2024awq, egiazarian2024extreme}. 
QuIP\# quantization used the QuIP-for-all framework~\cite{quip_for_all} to support Qwen-based architectures. All models were loaded via Hugging Face \texttt{transformers} to ensure a standardized inference framework.
%QuIP\# quantization used the QuIP-for-all framework\footnote{\url{https://github.com/chu-tianxiang/QuIP-for-all}} to support Qwen-based architectures. All models were loaded via Hugging Face \texttt{transformers} to ensure a standardized inference framework.

\subsection{Evaluation Metrics}
\label{sec:eval-metrics}

\noindent\textbf{Functional correctness.} We use pass@1~\cite{chen:arxiv2021}, which evaluates whether the model's top-ranked output passes all unit testsa strict single-attempt measure reflecting realistic deployment.

\smallskip
\noindent\textbf{Code quality.} We apply SonarCloud~\cite{sonarcloud} to assess five quality indicators. \emph{Reliability} quantifies code robustness by measuring bug density. \emph{Maintainability} captures code smells, suboptimal patterns that increase technical debt. \emph{Lines of Code (LoC)} counts non-whitespace lines as a proxy for implementation effort. \emph{Cyclomatic Complexity (CyC)} measures structural complexity via the control flow graph ($M = E + 2Q - N$, where $E$, $N$, and $Q$ denote edges, nodes, and connected components), with higher values indicating more branching paths. \emph{Cognitive Complexity (CoC)} captures human-perceived difficulty by accounting for nested control flow and conditional depth~\cite{munoz2020empirical}.

\subsection{Input Complexity Analysis (RQ$_3$)}
\label{sec:complexity-analysis}

To investigate whether prompt complexity modulates quantization-induced degradation, we analyze all Python tasks across McEval, CoderEval, and BigCodeBench. For each task, we compute \emph{token length} (word count) and \emph{Shannon entropy}~\cite{shannon1948mathematical} at the word level, capturing input size and lexical diversity respectively.

We pool all tasks and partition them into \textbf{High} and \textbf{Low} entropy buckets at the median, retaining benchmark identity within each bucket. Table~\ref{tab:entropy-buckets} summarizes the distribution: the High bucket is dominated by McEval and BigCodeBench (mean entropy $\approx$6.1), while the Low bucket is dominated by CoderEval (mean entropy $\approx$3.9). This partitioning captures benchmark-level complexity differences while enabling controlled within-bucket comparisons.

\begin{table}[t]
    \centering
    \caption{Distribution of tasks across entropy buckets. Tasks are pooled from McEval, CoderEval, and BigCodeBench (Python) and split at the median Shannon entropy.}
    \label{tab:entropy-buckets}
    \footnotesize
    \resizebox{\columnwidth}{!}{
    \begin{tabular}{llrrrr}
        \toprule
        \textbf{Bucket} & \textbf{Benchmark} & \textbf{Count} & \textbf{Mean Entropy} & \textbf{Std Entropy} & \textbf{Mean Length} \\
        \midrule
        \multirow{3}{*}{High}
        & McEval       & 30  & 6.08 & 0.18 & 145.60 \\
        & CoderEval    & 6   & 6.20 & 0.13 & 136.67 \\
        & BigCodeBench & 650 & 6.11 & 0.27 & 122.34 \\
        \midrule
        \multirow{3}{*}{Low}
        & McEval       & 12  & 5.30 & 0.39 & 79.92 \\
        & CoderEval    & 184 & 3.86 & 0.81 & 23.30 \\
        & BigCodeBench & 490 & 5.47 & 0.21 & 68.71 \\
        \bottomrule
    \end{tabular}
    }
    \vspace{-0.2cm}
\end{table}

We employ three complementary analyses: (1)~\textbf{McNemar's test}~\cite{mcnemar1947note} on 2$\times$2 contingency tables of discordant pairs to test whether quantization significantly alters per-task correctness; (2)~\textbf{stratified analysis} of pass@1 and degradation rates within each entropy bucket across the three benchmarks~\cite{cochran1954some}; and (3)~\textbf{point-biserial correlations}~\cite{cohen1988statistical} between prompt complexity (entropy, length) and degradation outcome, supplemented by Mann--Whitney U tests~\cite{mann1947test} comparing complexity distributions of degraded vs.\ non-degraded tasks.

\subsection{Analysis and Comparison Framework}
\label{sec:analysis}

We collect generated solutions from both Qwen2.5-Coder-7B-Instruct and CodeLlama-7B-Instruct across all McEval and CoderEval tasks in Python and Java (RQ$_1$, RQ$_2$), and across BigCodeBench tasks in Python (RQ$_3$). Each model is evaluated under seven configurations: one full-precision baseline and six 4-bit quantized variants. For every configuration, we capture the generated code and evaluate it along both dimensions: functional correctness via pass@1 and code quality via SonarCloud metrics.

To determine whether quantization introduces statistically significant changes, we conduct pairwise comparisons between each quantized variant and its full-precision counterpart. For pass@1 (binary), we apply \textbf{McNemar's test}~\cite{mcnemar1947note}, which evaluates whether the pattern of per-task pass/fail outcomes shifts significantly based on the 2$\times$2 contingency table of discordant pairs. For SonarCloud quality metrics (continuous), we apply the \textbf{Wilcoxon signed-rank test}~\cite{wilcoxon1945individual}, which assesses whether per-task quality score distributions differ significantly between full-precision and quantized variants. All p-values are adjusted via Holm--Bonferroni correction~\cite{holm1979simple} for six pairwise comparisons. Effect sizes are quantified using Cliff's delta~\cite{Cliff:2005}, categorized as negligible (N), small (S), medium (M), or large (L).
\rev{To account for output variability, we generated ten predictions per instance across all seven configurations of Qwen2.5-Coder-7B-Instruct on McEval-Python. Friedman's test~\cite{friedman1937use} found no significant run-level differences in pass@1 or any SonarCloud metric, indicating that the reported differences reflect technique effects rather than inference noise.}
% ============================================================
% Section 5: Results and Discussion
% ============================================================

\section{Results and Discussion}
\label{sec:results}
\begin{table*}[ht!]
    \centering
    \renewcommand{\arraystretch}{0.60}
    \setlength{\tabcolsep}{4pt}
    \caption{
Functional correctness (Pass@1) and code quality metrics of different models benchmarked on McEval-Java and CoderEval-Java. For quantized variants, {\color{green!60!black}$\blacktriangle$}~indicates improvement over the full-precision (FP) baseline, {\color{red}$\blacktriangledown$}~indicates degradation, and $\textbullet$~indicates no change. The highest Pass@1 per model--dataset pair is highlighted in \colorbox{green!15}{green}. CyC refers to Cyclomatic Complexity, while CoC denotes Cognitive Complexity.}
    \scriptsize
    \label{tab:model-performance-java}
    \resizebox{0.75\linewidth}{!}{%
        \begin{tabular}{llccc|cccccc}
            \toprule
            \textbf{Dataset} & \textbf{Model} & \textbf{Precision} & \textbf{PTQ Technique} & \textbf{Pass@1} & \multicolumn{6}{c}{\textbf{SonarCloud Metrics}} \\
            \cmidrule(lr){6-11}
            & & & & & \textbf{LoC} & \textbf{Security} & \textbf{Reliability} & \textbf{Maintainability} & \textbf{CyC} & \textbf{CoC} \\
            \midrule
            \multirow{14}{*}{\centering \emph{McEval-Java}} & \multirow{7}{*}{CodeLlama-7B} & \cellcolor[gray]{.75} 16 bit & \cellcolor[gray]{.75} \textbf{FP} & \cellcolor[gray]{.75} 0.25 & \cellcolor[gray]{.75} 1301 & \cellcolor[gray]{.75} 0 & \cellcolor[gray]{.75} 9 & \cellcolor[gray]{.75} 274 & \cellcolor[gray]{.75} 231 & \cellcolor[gray]{.75} 193 \\
            & & \multirow{6}{*}{4 bit} & AWQ & 0.23 & 1261$^{\color{green!60!black}\blacktriangle}$ & 0$^{\textbullet}$ & 9$^{\textbullet}$ & 328$^{\color{red}\blacktriangledown}$ & 229$^{\color{green!60!black}\blacktriangle}$ & 211$^{\color{red}\blacktriangledown}$ \\
            & &  & GPTQ & \cellcolor{green!15} 0.34 & 1384$^{\color{red}\blacktriangledown}$ & 0$^{\textbullet}$ & 9$^{\textbullet}$ & 276$^{\color{red}\blacktriangledown}$ & 243$^{\color{red}\blacktriangledown}$ & 202$^{\color{red}\blacktriangledown}$ \\
            & &  & GGUF & 0.23 & 1348$^{\color{red}\blacktriangledown}$ & 0$^{\textbullet}$ & 9$^{\textbullet}$ & 276$^{\color{red}\blacktriangledown}$ & 242$^{\color{red}\blacktriangledown}$ & 209$^{\color{red}\blacktriangledown}$ \\
            & &  & BitsAndBytes & 0.25 & 1368$^{\color{red}\blacktriangledown}$ & 0$^{\textbullet}$ & 9$^{\textbullet}$ & 280$^{\color{red}\blacktriangledown}$ & 251$^{\color{red}\blacktriangledown}$ & 251$^{\color{red}\blacktriangledown}$ \\
            & &  & AQLM & \cellcolor{green!15} 0.34 & 1418$^{\color{red}\blacktriangledown}$ & 0$^{\textbullet}$ & 9$^{\textbullet}$ & 274$^{\textbullet}$ & 269$^{\color{red}\blacktriangledown}$ & 273$^{\color{red}\blacktriangledown}$ \\
            & &  & QuIP\# & 0.08 & 1217$^{\color{green!60!black}\blacktriangle}$ & 0$^{\textbullet}$ & 10$^{\color{red}\blacktriangledown}$ & 250$^{\color{green!60!black}\blacktriangle}$ & 216$^{\color{green!60!black}\blacktriangle}$ & 193$^{\textbullet}$ \\
            \cline{2-11}
             & \multirow{7}{*}{Qwen2.5-Coder-7B} & \cellcolor[gray]{.75} 16 bit & \cellcolor[gray]{.75} \textbf{FP} & \cellcolor[gray]{.75} 0.45 & \cellcolor[gray]{.75} 1325 & \cellcolor[gray]{.75} 0 & \cellcolor[gray]{.75} 9 & \cellcolor[gray]{.75} 273 & \cellcolor[gray]{.75} 243 & \cellcolor[gray]{.75} 211 \\
            & & \multirow{6}{*}{4 bit} & AWQ & 0.45 & 1310$^{\color{green!60!black}\blacktriangle}$ & 0$^{\textbullet}$ & 9$^{\textbullet}$ & 328$^{\color{red}\blacktriangledown}$ & 252$^{\color{red}\blacktriangledown}$ & 234$^{\color{red}\blacktriangledown}$ \\
            & &  & GPTQ & 0.32 & 1222$^{\color{green!60!black}\blacktriangle}$ & 0$^{\textbullet}$ & 9$^{\textbullet}$ & 272$^{\color{green!60!black}\blacktriangle}$ & 215$^{\color{green!60!black}\blacktriangle}$ & 164$^{\color{green!60!black}\blacktriangle}$ \\
            & &  & GGUF & 0.51 & 1299$^{\color{green!60!black}\blacktriangle}$ & 0$^{\textbullet}$ & 9$^{\textbullet}$ & 278$^{\color{red}\blacktriangledown}$ & 234$^{\color{green!60!black}\blacktriangle}$ & 206$^{\color{green!60!black}\blacktriangle}$ \\
            & &  & BitsAndBytes & \cellcolor{green!15} 0.60 & 1368$^{\color{red}\blacktriangledown}$ & 0$^{\textbullet}$ & 9$^{\textbullet}$ & 275$^{\color{red}\blacktriangledown}$ & 261$^{\color{red}\blacktriangledown}$ & 252$^{\color{red}\blacktriangledown}$ \\
            & &  & AQLM & 0.58 & 1339$^{\color{red}\blacktriangledown}$ & 0$^{\textbullet}$ & 9$^{\textbullet}$ & 281$^{\color{red}\blacktriangledown}$ & 248$^{\color{red}\blacktriangledown}$ & 227$^{\color{red}\blacktriangledown}$ \\
            & &  & QuIP\# & 0.32 & 1237$^{\color{green!60!black}\blacktriangle}$ & 0$^{\textbullet}$ & 9$^{\textbullet}$ & 275$^{\color{red}\blacktriangledown}$ & 223$^{\color{green!60!black}\blacktriangle}$ & 203$^{\color{green!60!black}\blacktriangle}$ \\
            \midrule
            \multirow{14}{*}{\centering \emph{CoderEval-Java}} & \multirow{7}{*}{CodeLlama-7B} & \cellcolor[gray]{.75} 16 bit & \cellcolor[gray]{.75} \textbf{FP} & \cellcolor[gray]{.75} 0.31 & \cellcolor[gray]{.75} 1810 & \cellcolor[gray]{.75} 0 & \cellcolor[gray]{.75} 0 & \cellcolor[gray]{.75} 306 & \cellcolor[gray]{.75} 547 & \cellcolor[gray]{.75} 455 \\
            & & \multirow{6}{*}{4 bit} & AWQ & 0.28 & 1661$^{\color{green!60!black}\blacktriangle}$ & 0$^{\textbullet}$ & 0$^{\textbullet}$ & 290$^{\color{green!60!black}\blacktriangle}$ & 475$^{\color{green!60!black}\blacktriangle}$ & 369$^{\color{green!60!black}\blacktriangle}$ \\
            & &  & GPTQ & 0.30 & 1714$^{\color{green!60!black}\blacktriangle}$ & 0$^{\textbullet}$ & 1$^{\color{red}\blacktriangledown}$ & 311$^{\color{red}\blacktriangledown}$ & 474$^{\color{green!60!black}\blacktriangle}$ & 362$^{\color{green!60!black}\blacktriangle}$ \\
            & &  & GGUF & 0.28 & 1648$^{\color{green!60!black}\blacktriangle}$ & 0$^{\textbullet}$ & 0$^{\textbullet}$ & 286$^{\color{green!60!black}\blacktriangle}$ & 486$^{\color{green!60!black}\blacktriangle}$ & 386$^{\color{green!60!black}\blacktriangle}$ \\
            & &  & BitsAndBytes & 0.30 & 1706$^{\color{green!60!black}\blacktriangle}$ & 0$^{\textbullet}$ & 0$^{\textbullet}$ & 287$^{\color{green!60!black}\blacktriangle}$ & 534$^{\color{green!60!black}\blacktriangle}$ & 497$^{\color{red}\blacktriangledown}$ \\
            & &  & AQLM & \cellcolor{green!15} 0.33 & 1833$^{\color{red}\blacktriangledown}$ & 0$^{\textbullet}$ & 0$^{\textbullet}$ & 302$^{\color{green!60!black}\blacktriangle}$ & 535$^{\color{green!60!black}\blacktriangle}$ & 403$^{\color{green!60!black}\blacktriangle}$ \\
            & &  & QuIP\# & 0.28 & 2009$^{\color{red}\blacktriangledown}$ & 0$^{\textbullet}$ & 1$^{\color{red}\blacktriangledown}$ & 305$^{\color{green!60!black}\blacktriangle}$ & 603$^{\color{red}\blacktriangledown}$ & 573$^{\color{red}\blacktriangledown}$ \\
            \cline{2-11}
             & \multirow{7}{*}{Qwen2.5-Coder-7B} & \cellcolor[gray]{.75} 16 bit & \cellcolor[gray]{.75} \textbf{FP} & \cellcolor[gray]{.75} 0.20 & \cellcolor[gray]{.75} 1215 & \cellcolor[gray]{.75} 0 & \cellcolor[gray]{.75} 0 & \cellcolor[gray]{.75} 228 & \cellcolor[gray]{.75} 338 & \cellcolor[gray]{.75} 294 \\
            & & \multirow{6}{*}{4 bit} & AWQ & 0.20 & 1193$^{\color{green!60!black}\blacktriangle}$ & 0$^{\textbullet}$ & 0$^{\textbullet}$ & 201$^{\color{green!60!black}\blacktriangle}$ & 337$^{\color{green!60!black}\blacktriangle}$ & 270$^{\color{green!60!black}\blacktriangle}$ \\
            & &  & GPTQ & 0.18 & 1257$^{\color{red}\blacktriangledown}$ & 0$^{\textbullet}$ & 0$^{\textbullet}$ & 227$^{\color{green!60!black}\blacktriangle}$ & 352$^{\color{red}\blacktriangledown}$ & 335$^{\color{red}\blacktriangledown}$ \\
            & &  & GGUF & 0.24 & 1242$^{\color{red}\blacktriangledown}$ & 0$^{\textbullet}$ & 0$^{\textbullet}$ & 224$^{\color{green!60!black}\blacktriangle}$ & 344$^{\color{red}\blacktriangledown}$ & 239$^{\color{green!60!black}\blacktriangle}$ \\
            & &  & BitsAndBytes & 0.26 & 1293$^{\color{red}\blacktriangledown}$ & 0$^{\textbullet}$ & 0$^{\textbullet}$ & 250$^{\color{red}\blacktriangledown}$ & 355$^{\color{red}\blacktriangledown}$ & 319$^{\color{red}\blacktriangledown}$ \\
            & &  & AQLM & \cellcolor{green!15} 0.27 & 1211$^{\color{green!60!black}\blacktriangle}$ & 0$^{\textbullet}$ & 0$^{\textbullet}$ & 250$^{\color{red}\blacktriangledown}$ & 356$^{\color{red}\blacktriangledown}$ & 289$^{\color{green!60!black}\blacktriangle}$ \\
            & &  & QuIP\# & 0.16 & 1049$^{\color{green!60!black}\blacktriangle}$ & 0$^{\textbullet}$ & 0$^{\textbullet}$ & 194$^{\color{green!60!black}\blacktriangle}$ & 302$^{\color{green!60!black}\blacktriangle}$ & 225$^{\color{green!60!black}\blacktriangle}$ \\
            \bottomrule
        \end{tabular}%
    }
    %\vspace{-0.2cm}
\end{table*}

We present and discuss the results of our empirical study, organized by research question.

% ----------------------------------------------------------
\smallskip
\noindent\textbf{RQ$_1$: How do different quantization techniques impact the functional correctness of code generated by large code models?}

\smallskip
Table~\ref{tab:model-performance-java} presents the pass@1 scores for both model families on McEval-Java and CoderEval-Java. On McEval-Java, CodeLlama-7B (FP: 0.25) exhibits wide variation across techniques: GPTQ and AQLM both reach 0.34, while QuIP\# drops to 0.08---a substantial reduction indicating that this rotation-based approach is particularly detrimental for CodeLlama on this benchmark. AWQ (0.23) and GGUF (0.23) show marginal degradation, while BitsAndBytes preserves the baseline exactly (0.25). For Qwen2.5-Coder-7B (FP: 0.45), the pattern differs considerably: BitsAndBytes achieves the highest pass@1 at 0.60---a gain of 15 percentage points---followed by AQLM (0.58) and GGUF (0.51). AWQ matches the baseline (0.45), while GPTQ (0.32) and QuIP\# (0.32) exhibit meaningful losses.

On CoderEval-Java---which features pragmatic code generation tasks with real-world context dependencies---the quantized variants cluster more tightly around their baselines. For CodeLlama-7B (FP: 0.31), AQLM (0.33) slightly exceeds FP, while the remaining techniques produce pass@1 scores between 0.28 and 0.30. Even QuIP\# (0.28), which suffered substantially on McEval, shows only modest decline---suggesting its degradation pattern is benchmark-sensitive rather than uniform. For Qwen2.5-Coder-7B (FP: 0.20), AQLM leads at 0.27, followed by BitsAndBytes (0.26) and GGUF (0.24).

Two cross-cutting observations emerge. First, the impact of quantization is highly technique-dependent: within the same model and benchmark, pass@1 can range from substantial degradation (QuIP\# on CodeLlama-McEval) to notable improvement (BitsAndBytes on Qwen-McEval). Second, the effect is model-dependent: techniques that degrade one model may benefit another. For instance, GPTQ improves CodeLlama's McEval pass@1 by 9 percentage points but reduces Qwen's by 13.

McNemar's test (detailed statistical tables are available in our online appendix~\cite{replication}) confirms that only one Java comparison reaches significance: CodeLlama FP vs.\ QuIP\# on McEval ($p < 0.05$, OR = 19.00). All other pairwise comparisons are non-significant, indicating that despite considerable variation in aggregate scores, task-level correctness patterns are largely preserved under 4-bit quantization.

Table~\ref{tab:model-performance-python} presents the Python results across BCB-Python, McEval-Python, and CoderEval-Python. The patterns largely mirror Java. On McEval-Python, BitsAndBytes achieves the highest pass@1 for CodeLlama (0.19 vs.\ FP 0.17), while GGUF and AQLM both reach 0.43 for Qwen (vs.\ FP 0.31). On CoderEval-Python, AQLM leads for both models (CodeLlama: 0.26; Qwen: 0.31). The BCB-Python benchmark reveals the starkest contrasts: AQLM achieves the best CodeLlama pass@1 (0.27 vs.\ FP 0.23), while QuIP\# drops to 0.12---roughly halving the baseline. Qwen's quantized variants preserve correctness more consistently, with AWQ, GPTQ, and AQLM all matching FP (0.41); QuIP\# again shows the largest drop (0.32).

The Python statistical analysis reveals more significant findings than Java, concentrated on BCB-Python: CodeLlama QuIP\# ($p < 0.05$, OR = 5.10) and BitsAndBytes ($p < 0.05$, OR = 2.05) significantly degrade correctness, while AQLM ($p < 0.05$, OR = 0.62) significantly \emph{improves} it. For Qwen, only QuIP\# reaches significance ($p < 0.05$, OR = 2.71). McEval-Python and CoderEval-Python comparisons remain non-significant. Across both languages, AQLM is the most consistently competitive variant, frequently matching or exceeding FP, while QuIP\# exhibits the widest performance variance. Statistical significance is concentrated on BCB-Python, suggesting that quantization-induced correctness shifts become more detectable as task complexity increases---a pattern we examine in RQ$_3$.

% \nopagebreak
% \begin{tcolorbox}[colback=gray!5, colframe=black!80, title=Summary -- RQ$_1$]
% At 4-bit precision, quantization largely preserves functional correctness across both languages. AQLM consistently matches or exceeds the FP baseline, while QuIP\# is the most degradation-prone---particularly on complex tasks (BCB-Python), where it reaches significance for both models. The remaining techniques cluster close to their baselines, with no significant correctness shifts on simpler benchmarks.
% \end{tcolorbox}
% \newtcolorbox{findingbox}{
%   enhanced, breakable,
%   colback=gray!10!white,
%   colframe=gray!10!white,
%   arc=8pt, boxrule=0pt,
%   boxsep=3pt, left=10pt, right=10pt,
%   top=6pt, bottom=6pt
% }

%\FloatBarrier
\smallskip
\nopagebreak 
\begin{tcolorbox}[summarybox, title=\textbf{Summary -- RQ$_1$}]
At 4-bit precision, quantization largely preserves functional correctness across both languages. AQLM consistently matches or exceeds the FP baseline, while QuIP\# is the most degradation-prone---particularly on complex tasks (BCB-Python), where it reaches significance for both models. The remaining techniques cluster close to their baselines, with no significant correctness shifts on simpler benchmarks.
\end{tcolorbox}

\smallskip
\nopagebreak

\smallskip
\noindent\textbf{RQ$_2$: How do different quantization techniques impact the quality attributes of automatically generated code?}

\begin{table*}[ht!]
    \centering
    \renewcommand{\arraystretch}{0.90}
    \setlength{\tabcolsep}{6pt}
    \caption{
Functional correctness (Pass@1) and code quality metrics of different models benchmarked on BCB-Python, McEval-Python, and CoderEval-Python. For quantized variants, {\color{green!60!black}$\blacktriangle$}~indicates improvement over the full-precision (FP) baseline, {\color{red}$\blacktriangledown$}~indicates degradation, and $\textbullet$~indicates no change. The highest Pass@1 per model--dataset pair is highlighted in \colorbox{green!15}{green}. CyC refers to Cyclomatic Complexity, while CoC denotes Cognitive Complexity.}
    \footnotesize
    \label{tab:model-performance-python}
    \resizebox{0.80\linewidth}{!}{%
        \begin{tabular}{llccc|cccccc}
            \toprule
            \textbf{Dataset} & \textbf{Model} & \textbf{Precision} & \textbf{PTQ Technique} & \textbf{Pass@1} & \multicolumn{6}{c}{\textbf{SonarCloud Metrics}} \\
            \cmidrule(lr){6-11}
            & & & & & \textbf{LoC} & \textbf{Security} & \textbf{Reliability} & \textbf{Maintainability} & \textbf{CyC} & \textbf{CoC} \\
            \midrule
            \multirow{14}{*}{\centering \emph{BCB-Python}} & \multirow{7}{*}{CodeLlama-7B} & \cellcolor[gray]{.75} 16 bit & \cellcolor[gray]{.75} \textbf{FP} & \cellcolor[gray]{.75} 0.23 & \cellcolor[gray]{.75} 17484 & \cellcolor[gray]{.75} 0 & \cellcolor[gray]{.75} 22 & \cellcolor[gray]{.75} 819 & \cellcolor[gray]{.75} 2726 & \cellcolor[gray]{.75} 2109 \\
            & & \multirow{6}{*}{4 bit} & AWQ & 0.22 & 17414$^{\color{green!60!black}\blacktriangle}$ & 0$^{\textbullet}$ & 27$^{\color{red}\blacktriangledown}$ & 788$^{\color{green!60!black}\blacktriangle}$ & 2848$^{\color{red}\blacktriangledown}$ & 2273$^{\color{red}\blacktriangledown}$ \\
            & &  & GPTQ & 0.21 & 17776$^{\color{red}\blacktriangledown}$ & 0$^{\textbullet}$ & 17$^{\color{green!60!black}\blacktriangle}$ & 804$^{\color{green!60!black}\blacktriangle}$ & 2832$^{\color{red}\blacktriangledown}$ & 2170$^{\color{red}\blacktriangledown}$ \\
            & &  & GGUF & 0.21 & 15623$^{\color{green!60!black}\blacktriangle}$ & 0$^{\textbullet}$ & 17$^{\color{green!60!black}\blacktriangle}$ & 735$^{\color{green!60!black}\blacktriangle}$ & 2675$^{\color{green!60!black}\blacktriangle}$ & 1988$^{\color{green!60!black}\blacktriangle}$ \\
            & &  & BitsAndBytes & 0.18 & 16070$^{\color{green!60!black}\blacktriangle}$ & 0$^{\textbullet}$ & 15$^{\color{green!60!black}\blacktriangle}$ & 782$^{\color{green!60!black}\blacktriangle}$ & 2688$^{\color{green!60!black}\blacktriangle}$ & 2051$^{\color{green!60!black}\blacktriangle}$ \\
            & &  & AQLM & \cellcolor{green!15} 0.27 & 17006$^{\color{green!60!black}\blacktriangle}$ & 0$^{\textbullet}$ & 32$^{\color{red}\blacktriangledown}$ & 883$^{\color{red}\blacktriangledown}$ & 3179$^{\color{red}\blacktriangledown}$ & 2446$^{\color{red}\blacktriangledown}$ \\
            & &  & QuIP\# & 0.12 & 15703$^{\color{green!60!black}\blacktriangle}$ & 0$^{\textbullet}$ & 43$^{\color{red}\blacktriangledown}$ & 886$^{\color{red}\blacktriangledown}$ & 2641$^{\color{green!60!black}\blacktriangle}$ & 1883$^{\color{green!60!black}\blacktriangle}$ \\
            \cline{2-11}
             & \multirow{7}{*}{Qwen2.5-Coder-7B} & \cellcolor[gray]{.75} 16 bit & \cellcolor[gray]{.75} \textbf{FP} & \cellcolor{green!15} 0.41 & \cellcolor[gray]{.75} 17281 & \cellcolor[gray]{.75} 0 & \cellcolor[gray]{.75} 21 & \cellcolor[gray]{.75} 670 & \cellcolor[gray]{.75} 2975 & \cellcolor[gray]{.75} 2492 \\
            & & \multirow{6}{*}{4 bit} & AWQ & 0.41 & 17184$^{\color{green!60!black}\blacktriangle}$ & 0$^{\textbullet}$ & 12$^{\color{green!60!black}\blacktriangle}$ & 631$^{\color{green!60!black}\blacktriangle}$ & 2905$^{\color{green!60!black}\blacktriangle}$ & 2409$^{\color{green!60!black}\blacktriangle}$ \\
            & &  & GPTQ & 0.41 & 17320$^{\color{red}\blacktriangledown}$ & 0$^{\textbullet}$ & 11$^{\color{green!60!black}\blacktriangle}$ & 679$^{\color{red}\blacktriangledown}$ & 2927$^{\color{green!60!black}\blacktriangle}$ & 2458$^{\color{green!60!black}\blacktriangle}$ \\
            & &  & GGUF & 0.40 & 17148$^{\color{green!60!black}\blacktriangle}$ & 0$^{\textbullet}$ & 22$^{\color{red}\blacktriangledown}$ & 664$^{\color{green!60!black}\blacktriangle}$ & 2983$^{\color{red}\blacktriangledown}$ & 2565$^{\color{red}\blacktriangledown}$ \\
            & &  & BitsAndBytes & 0.39 & 17909$^{\color{red}\blacktriangledown}$ & 0$^{\textbullet}$ & 30$^{\color{red}\blacktriangledown}$ & 771$^{\color{red}\blacktriangledown}$ & 3088$^{\color{red}\blacktriangledown}$ & 2718$^{\color{red}\blacktriangledown}$ \\
            & &  & AQLM & 0.41 & 17278$^{\color{green!60!black}\blacktriangle}$ & 0$^{\textbullet}$ & 19$^{\color{green!60!black}\blacktriangle}$ & 667$^{\color{green!60!black}\blacktriangle}$ & 2987$^{\color{red}\blacktriangledown}$ & 2577$^{\color{red}\blacktriangledown}$ \\
            & &  & QuIP\# & 0.32 & 16547$^{\color{green!60!black}\blacktriangle}$ & 0$^{\textbullet}$ & 14$^{\color{green!60!black}\blacktriangle}$ & 733$^{\color{red}\blacktriangledown}$ & 2841$^{\color{green!60!black}\blacktriangle}$ & 2207$^{\color{green!60!black}\blacktriangle}$ \\
            \midrule
            \multirow{14}{*}{\centering \emph{McEval-Python}} & \multirow{7}{*}{CodeLlama-7B} & \cellcolor[gray]{.75} 16 bit & \cellcolor[gray]{.75} \textbf{FP} & \cellcolor[gray]{.75} 0.17 & \cellcolor[gray]{.75} 1025 & \cellcolor[gray]{.75} 0 & \cellcolor[gray]{.75} 1 & \cellcolor[gray]{.75} 64 & \cellcolor[gray]{.75} 207 & \cellcolor[gray]{.75} 215 \\
            & & \multirow{6}{*}{4 bit} & AWQ & 0.17 & 977$^{\color{green!60!black}\blacktriangle}$ & 0$^{\textbullet}$ & 2$^{\color{red}\blacktriangledown}$ & 32$^{\color{green!60!black}\blacktriangle}$ & 201$^{\color{green!60!black}\blacktriangle}$ & 177$^{\color{green!60!black}\blacktriangle}$ \\
            & &  & GPTQ & 0.14 & 1007$^{\color{green!60!black}\blacktriangle}$ & 0$^{\textbullet}$ & 0$^{\color{green!60!black}\blacktriangle}$ & 88$^{\color{red}\blacktriangledown}$ & 215$^{\color{red}\blacktriangledown}$ & 226$^{\color{red}\blacktriangledown}$ \\
            & &  & GGUF & 0.12 & 949$^{\color{green!60!black}\blacktriangle}$ & 0$^{\textbullet}$ & 0$^{\color{green!60!black}\blacktriangle}$ & 26$^{\color{green!60!black}\blacktriangle}$ & 194$^{\color{green!60!black}\blacktriangle}$ & 187$^{\color{green!60!black}\blacktriangle}$ \\
            & &  & BitsAndBytes & \cellcolor{green!15} 0.19 & 925$^{\color{green!60!black}\blacktriangle}$ & 0$^{\textbullet}$ & 2$^{\color{red}\blacktriangledown}$ & 51$^{\color{green!60!black}\blacktriangle}$ & 187$^{\color{green!60!black}\blacktriangle}$ & 168$^{\color{green!60!black}\blacktriangle}$ \\
            & &  & AQLM & 0.12 & 860$^{\color{green!60!black}\blacktriangle}$ & 0$^{\textbullet}$ & 0$^{\color{green!60!black}\blacktriangle}$ & 51$^{\color{green!60!black}\blacktriangle}$ & 194$^{\color{green!60!black}\blacktriangle}$ & 138$^{\color{green!60!black}\blacktriangle}$ \\
            & &  & QuIP\# & 0.10 & 1006$^{\color{green!60!black}\blacktriangle}$ & 0$^{\textbullet}$ & 2$^{\color{red}\blacktriangledown}$ & 80$^{\color{red}\blacktriangledown}$ & 198$^{\color{green!60!black}\blacktriangle}$ & 194$^{\color{green!60!black}\blacktriangle}$ \\
            \cline{2-11}
             & \multirow{7}{*}{Qwen2.5-Coder-7B} & \cellcolor[gray]{.75} 16 bit & \cellcolor[gray]{.75} \textbf{FP} & \cellcolor[gray]{.75} 0.31 & \cellcolor[gray]{.75} 816 & \cellcolor[gray]{.75} 0 & \cellcolor[gray]{.75} 0 & \cellcolor[gray]{.75} 25 & \cellcolor[gray]{.75} 150 & \cellcolor[gray]{.75} 107 \\
            & & \multirow{6}{*}{4 bit} & AWQ & 0.24 & 817$^{\color{red}\blacktriangledown}$ & 0$^{\textbullet}$ & 0$^{\textbullet}$ & 26$^{\color{red}\blacktriangledown}$ & 157$^{\color{red}\blacktriangledown}$ & 106$^{\color{green!60!black}\blacktriangle}$ \\
            & &  & GPTQ & 0.19 & 779$^{\color{green!60!black}\blacktriangle}$ & 0$^{\textbullet}$ & 0$^{\textbullet}$ & 30$^{\color{red}\blacktriangledown}$ & 134$^{\color{green!60!black}\blacktriangle}$ & 80$^{\color{green!60!black}\blacktriangle}$ \\
            & &  & GGUF & \cellcolor{green!15} 0.43 & 953$^{\color{red}\blacktriangledown}$ & 0$^{\textbullet}$ & 0$^{\textbullet}$ & 31$^{\color{red}\blacktriangledown}$ & 205$^{\color{red}\blacktriangledown}$ & 193$^{\color{red}\blacktriangledown}$ \\
            & &  & BitsAndBytes & 0.38 & 984$^{\color{red}\blacktriangledown}$ & 0$^{\textbullet}$ & 1$^{\color{red}\blacktriangledown}$ & 25$^{\textbullet}$ & 223$^{\color{red}\blacktriangledown}$ & 224$^{\color{red}\blacktriangledown}$ \\
            & &  & AQLM & \cellcolor{green!15} 0.43 & 958$^{\color{red}\blacktriangledown}$ & 0$^{\textbullet}$ & 1$^{\color{red}\blacktriangledown}$ & 33$^{\color{red}\blacktriangledown}$ & 210$^{\color{red}\blacktriangledown}$ & 212$^{\color{red}\blacktriangledown}$ \\
            & &  & QuIP\# & 0.36 & 944$^{\color{red}\blacktriangledown}$ & 0$^{\textbullet}$ & 1$^{\color{red}\blacktriangledown}$ & 24$^{\color{green!60!black}\blacktriangle}$ & 194$^{\color{red}\blacktriangledown}$ & 157$^{\color{red}\blacktriangledown}$ \\
            \midrule
            \multirow{14}{*}{\centering \emph{CoderEval-Python}} & \multirow{7}{*}{CodeLlama-7B} & \cellcolor[gray]{.75} 16 bit & \cellcolor[gray]{.75} \textbf{FP} & \cellcolor[gray]{.75} 0.24 & \cellcolor[gray]{.75} 1378 & \cellcolor[gray]{.75} 0 & \cellcolor[gray]{.75} 24 & \cellcolor[gray]{.75} 100 & \cellcolor[gray]{.75} 489 & \cellcolor[gray]{.75} 507 \\
            & & \multirow{6}{*}{4 bit} & AWQ & 0.24 & 1255$^{\color{green!60!black}\blacktriangle}$ & 0$^{\textbullet}$ & 5$^{\color{green!60!black}\blacktriangle}$ & 71$^{\color{green!60!black}\blacktriangle}$ & 420$^{\color{green!60!black}\blacktriangle}$ & 392$^{\color{green!60!black}\blacktriangle}$ \\
            & &  & GPTQ & 0.23 & 1316$^{\color{green!60!black}\blacktriangle}$ & 0$^{\textbullet}$ & 4$^{\color{green!60!black}\blacktriangle}$ & 47$^{\color{green!60!black}\blacktriangle}$ & 442$^{\color{green!60!black}\blacktriangle}$ & 424$^{\color{green!60!black}\blacktriangle}$ \\
            & &  & GGUF & 0.23 & 1365$^{\color{green!60!black}\blacktriangle}$ & 0$^{\textbullet}$ & 15$^{\color{green!60!black}\blacktriangle}$ & 102$^{\color{red}\blacktriangledown}$ & 475$^{\color{green!60!black}\blacktriangle}$ & 522$^{\color{red}\blacktriangledown}$ \\
            & &  & BitsAndBytes & 0.22 & 1195$^{\color{green!60!black}\blacktriangle}$ & 0$^{\textbullet}$ & 1$^{\color{green!60!black}\blacktriangle}$ & 91$^{\color{green!60!black}\blacktriangle}$ & 432$^{\color{green!60!black}\blacktriangle}$ & 400$^{\color{green!60!black}\blacktriangle}$ \\
            & &  & AQLM & \cellcolor{green!15} 0.26 & 1460$^{\color{red}\blacktriangledown}$ & 0$^{\textbullet}$ & 4$^{\color{green!60!black}\blacktriangle}$ & 142$^{\color{red}\blacktriangledown}$ & 526$^{\color{red}\blacktriangledown}$ & 560$^{\color{red}\blacktriangledown}$ \\
            & &  & QuIP\# & 0.21 & 1138$^{\color{green!60!black}\blacktriangle}$ & 0$^{\textbullet}$ & 9$^{\color{green!60!black}\blacktriangle}$ & 52$^{\color{green!60!black}\blacktriangle}$ & 433$^{\color{green!60!black}\blacktriangle}$ & 416$^{\color{green!60!black}\blacktriangle}$ \\
            \cline{2-11}
             & \multirow{7}{*}{Qwen2.5-Coder-7B} & \cellcolor[gray]{.75} 16 bit & \cellcolor[gray]{.75} \textbf{FP} & \cellcolor[gray]{.75} 0.27 & \cellcolor[gray]{.75} 988 & \cellcolor[gray]{.75} 0 & \cellcolor[gray]{.75} 1 & \cellcolor[gray]{.75} 39 & \cellcolor[gray]{.75} 350 & \cellcolor[gray]{.75} 315 \\
            & & \multirow{6}{*}{4 bit} & AWQ & 0.28 & 1500$^{\color{red}\blacktriangledown}$ & 0$^{\textbullet}$ & 23$^{\color{red}\blacktriangledown}$ & 36$^{\color{green!60!black}\blacktriangle}$ & 500$^{\color{red}\blacktriangledown}$ & 495$^{\color{red}\blacktriangledown}$ \\
            & &  & GPTQ & 0.23 & 1487$^{\color{red}\blacktriangledown}$ & 0$^{\textbullet}$ & 32$^{\color{red}\blacktriangledown}$ & 41$^{\color{red}\blacktriangledown}$ & 488$^{\color{red}\blacktriangledown}$ & 515$^{\color{red}\blacktriangledown}$ \\
            & &  & GGUF & 0.26 & 1541$^{\color{red}\blacktriangledown}$ & 0$^{\textbullet}$ & 12$^{\color{red}\blacktriangledown}$ & 33$^{\color{green!60!black}\blacktriangle}$ & 524$^{\color{red}\blacktriangledown}$ & 540$^{\color{red}\blacktriangledown}$ \\
            & &  & BitsAndBytes & 0.27 & 1618$^{\color{red}\blacktriangledown}$ & 0$^{\textbullet}$ & 23$^{\color{red}\blacktriangledown}$ & 47$^{\color{red}\blacktriangledown}$ & 526$^{\color{red}\blacktriangledown}$ & 533$^{\color{red}\blacktriangledown}$ \\
            & &  & AQLM & \cellcolor{green!15} 0.31 & 1469$^{\color{red}\blacktriangledown}$ & 0$^{\textbullet}$ & 5$^{\color{red}\blacktriangledown}$ & 28$^{\color{green!60!black}\blacktriangle}$ & 509$^{\color{red}\blacktriangledown}$ & 498$^{\color{red}\blacktriangledown}$ \\
            & &  & QuIP\# & 0.24 & 1367$^{\color{red}\blacktriangledown}$ & 0$^{\textbullet}$ & 23$^{\color{red}\blacktriangledown}$ & 48$^{\color{red}\blacktriangledown}$ & 461$^{\color{red}\blacktriangledown}$ & 428$^{\color{red}\blacktriangledown}$ \\
            \bottomrule
        \end{tabular}
    }
    %\vspace{-0.2cm}
\end{table*}

\smallskip
Tables~\ref{tab:model-performance-java} and~\ref{tab:model-performance-python} report SonarCloud metrics---LoC, Security, Reliability, Maintainability, Cyclomatic Complexity (CyC), and Cognitive Complexity (CoC)---for all configurations.

Beginning with McEval-Java, we observe a mixed pattern for CodeLlama-7B. AWQ and QuIP\# reduce LoC relative to the baseline (1,261 and 1,217 vs.\ 1,301), while AQLM produces the longest code (1,418). Security remains unchanged across all techniques (0 hotspots), and Reliability is stable at 9 for all variants except QuIP\#, which introduces one additional issue. Maintainability shows notable variation: AWQ increases code smells substantially (328 vs.\ 274 in FP), while QuIP\# reduces them (250). For Qwen2.5-Coder-7B, GPTQ stands out as the most quality-preserving technique, improving LoC (1,222 vs.\ 1,325), CyC (215 vs.\ 243), and CoC (164 vs.\ 211). In contrast, AWQ again inflates Maintainability (328 vs.\ 273)---a pattern consistent across both model families.

On CoderEval-Java, most CodeLlama variants produce \emph{fewer} quality issues than FP---AWQ, GGUF, and BitsAndBytes all reduce LoC, Maintainability, and CyC. The exceptions are AQLM and QuIP\#, which generate longer code (1,833 and 2,009 vs.\ 1,810) with higher complexity. For Qwen, QuIP\# produces the most compact code across all metrics (LoC: 1,049, Maintainability: 194, CyC: 302, CoC: 225), while BitsAndBytes and AQLM tend to increase complexity. Directional improvements from quantized models likely reflect stochastic variation rather than systematic quality gains.

The Wilcoxon signed-rank test (see online appendix~\cite{replication}) confirms that most quality differences are non-significant. The clearest finding is AWQ's effect on Maintainability: both CodeLlama ($d = -0.346$, medium) and Qwen ($d = -0.394$, medium) show significant increases in code smells on McEval-Java, suggesting AWQ's activation-aware scaling systematically alters patterns flagged by static analysis. On CoderEval-Java, AWQ shows significant LoC, CyC, and CoC differences for CodeLlama, though all with negligible effect sizes. Across all Java comparisons, Security and Reliability remain unchanged, confirming that quantization does not introduce security vulnerabilities or reliability regressions.

Turning to Python, the quality landscape is more differentiated. On McEval-Python, CodeLlama exhibits a pattern reminiscent of Java: most variants reduce LoC relative to FP, Security remains zero, and directional changes in Maintainability are mixed. For Qwen, BitsAndBytes substantially increases complexity (CyC: 223 vs.\ 150; CoC: 224 vs.\ 107), while GPTQ reduces both (CyC: 134; CoC: 80). On CoderEval-Python, a notable asymmetry emerges: CodeLlama's quantized variants generally improve quality relative to FP, while \emph{all six} Qwen variants substantially increase LoC (1,367--1,618 vs.\ 988), CyC, and CoC---a consistent quality degradation unique to this model--benchmark--language combination. On BCB-Python, QuIP\# inflates CodeLlama's Reliability (43 vs.\ 22) and Maintainability (886 vs.\ 819) despite reducing LoC; BitsAndBytes shows the largest divergence for Qwen across LoC (17,909 vs.\ 17,281), Reliability (30 vs.\ 21), and complexity.

The Python statistical analysis reveals substantially more significant findings. On CoderEval-Python, all six Qwen variants show significant LoC differences (small to medium effects), with most also showing significant CyC and CoC shifts---confirming that this quality degradation is statistically robust. On McEval-Python, BitsAndBytes produces the only \emph{large} effect size in our entire study: CyC ($d = -0.452$, medium) and CoC ($d = -0.487$, large) for Qwen. On BCB-Python, the large sample size enables detection of many significant but overwhelmingly negligible effects. Security remains zero across all Python configurations. Synthesizing across languages, Security is universally unaffected, Maintainability is the most sensitive dimension (AWQ in Java, BitsAndBytes in Python), and Java quality differences are largely non-significant while Python---especially CoderEval and BCB for Qwen---reveals statistically robust quality shifts.

% attach boxed title to top left={yshift=-0pt},
% boxed title style={
%   colback=navy, arc=3pt 3pt 0pt 0pt,
%   boxrule=0pt, colframe=navy,
%   fontupper=\scriptsize\bfseries\color{white}
% },
% colback=blue!6!white,
% colframe=blue!30!black!40,
% arc=0pt 6pt 6pt 6pt,
% boxrule=0.4pt

% \smallskip
% \nopagebreak
% \begin{tcolorbox}[summarybox, title=\textbf{Summary -- RQ$_2$}]
% 4-bit quantization broadly preserves code quality; Security remains unaffected. AWQ raises Maintainability issues in Java (medium effect) and BitsAndBytes drives the largest quality shift in Python (large effect), with Qwen2.5-Coder-7B on CoderEval-Python most sensitive overall.
%\end{tcolorbox}
% \smallskip

\smallskip
\nopagebreak
\begin{tcolorbox}[summarybox, title=\textbf{Summary -- RQ$_2$}]
Quantization at 4-bit generally preserves code quality, with Security unaffected and most effects negligible. AWQ consistently increases Maintainability issues in Java (medium effect), while BitsAndBytes introduces the largest observed effect in Python (CyC/CoC, large effect). CoderEval-Python Qwen2.5-Coder-7B is the most quality-sensitive configuration, with all six techniques showing significant LoC and complexity increases.
\end{tcolorbox}
\smallskip

\begin{table*}[!t]
    \centering
    \renewcommand{\arraystretch}{0.70}
    \setlength{\tabcolsep}{4pt}
    \caption{
Per-technique degradation rates (\%) within entropy buckets and point-biserial correlation between input complexity and quantization-induced degradation. Tasks from McEval, CoderEval, and BigCodeBench (Python) are pooled and split at the median Shannon entropy. Correlations ($r$) are computed over all pooled tasks. Significant correlations ($p < 0.05$) are highlighted in \colorbox{red!10}{red}. {*}$p<0.05$, {**}$p<0.01$, {***}$p<0.001$.}
    \label{tab:rq3-combined}
    %\footnotesize
    \scriptsize
    %\resizebox{\linewidth}{!}{
    \begin{tabular}{ll rrr rrr rl rl}
        \toprule
        & & \multicolumn{3}{c}{\textbf{Degr.\ Rate on High-Entropy (\%)}} & \multicolumn{3}{c}{\textbf{Degr.\ Rate on Low-Entropy (\%)}} & \multicolumn{2}{c}{\textbf{Entropy Corr.}} & \multicolumn{2}{c}{\textbf{Length Corr.}} \\
        \cmidrule(lr){3-5} \cmidrule(lr){6-8} \cmidrule(lr){9-10} \cmidrule(lr){11-12}
        \textbf{Model} & \textbf{PTQ Technique} & \textbf{McE} & \textbf{CodE} & \textbf{BCB} & \textbf{McE} & \textbf{CodE} & \textbf{BCB} & \textbf{$r$} & \textbf{$p$} & \textbf{$r$} & \textbf{$p$} \\
        \midrule
        \multirow{6}{*}{CodeLlama-7B}
         & AWQ          & 6.7           & 0.0  & 5.5           & 0.0  & 2.2  & 4.7           & \cellcolor{red!10}$+$0.192 & \cellcolor{red!10}*** & \cellcolor{red!10}$+$0.248 & \cellcolor{red!10}*** \\
         & GPTQ         & 6.7           & 0.0  & 4.3           & 0.0  & 3.8  & 4.7           & $+$0.092 & 0.108 & $+$0.106 & 0.063 \\
         & GGUF         & 10.0          & 0.0  & 5.4           & 0.0  & 3.8  & 8.6           & \cellcolor{red!10}$+$0.137 & \cellcolor{red!10}* & \cellcolor{red!10}$+$0.207 & \cellcolor{red!10}*** \\
         & BitsAndBytes & 3.3           & 0.0  & 7.4           & 0.0  & 2.7  & 11.6          & \cellcolor{red!10}$+$0.203 & \cellcolor{red!10}*** & \cellcolor{red!10}$+$0.187 & \cellcolor{red!10}*** \\
         & AQLM         & \textbf{16.7} & 0.0  & 6.9           & 0.0  & 3.8  & 9.4           & \cellcolor{red!10}$+$0.177 & \cellcolor{red!10}** & \cellcolor{red!10}$+$0.172 & \cellcolor{red!10}** \\
         & QuIP\#       & 13.3          & 0.0  & \textbf{12.3} & 0.0  & \textbf{7.1} & \textbf{15.3} & \cellcolor{red!10}$+$0.325 & \cellcolor{red!10}*** & \cellcolor{red!10}$+$0.321 & \cellcolor{red!10}*** \\
        \midrule
        \multirow{6}{*}{Qwen2.5-Coder-7B}
         & AWQ          & 6.7           & 0.0  & 5.2           & \textbf{16.7} & 5.4  & 4.9  & $-$0.039 & 0.374 & $-$0.004 & 0.923 \\
         & GPTQ         & \textbf{13.3} & 0.0  & 3.7           & 8.3  & \textbf{8.2} & 5.9  & \cellcolor{red!10}$-$0.156 & \cellcolor{red!10}*** & \cellcolor{red!10}$-$0.086 & \cellcolor{red!10}* \\
         & GGUF         & 3.3           & 0.0  & 5.1           & 8.3  & 4.9  & 5.7           & $-$0.049 & 0.263 & $-$0.029 & 0.506 \\
         & BitsAndBytes & 6.7           & 0.0  & 7.8           & \textbf{16.7} & 5.4  & 9.8  & $-$0.011 & 0.807 & $+$0.015 & 0.734 \\
         & AQLM         & 6.7           & 0.0  & 6.3           & 8.3  & 4.3  & 7.5           & $+$0.014 & 0.741 & $+$0.054 & 0.213 \\
         & QuIP\#       & 6.7           & \textbf{16.7} & \textbf{14.8} & \textbf{16.7} & 7.6 & \textbf{15.9} & $+$0.061 & 0.160 & \cellcolor{red!10}$+$0.095 & \cellcolor{red!10}* \\
        \bottomrule
    \end{tabular}
    %}
    %\vspace{-0.2cm}
\end{table*}
\smallskip
\noindent\textbf{RQ$_3$: Does input prompt complexity influence quantization-induced correctness degradation?}

\smallskip
We pool all Python tasks from McEval, CoderEval, and BigCodeBench, split them at the median Shannon entropy into High and Low complexity buckets (Table~\ref{tab:entropy-buckets}), and measure the point-biserial correlation between complexity (entropy and token length) and the binary degradation outcome---\ie whether a task solved by FP was broken by the quantized variant. Table~\ref{tab:rq3-combined} presents per-benchmark degradation rates within each bucket alongside the correlation results.

Across both models, CoderEval tasks---short docstring-style prompts dominating the Low bucket---exhibit near-zero degradation. BigCodeBench concentrates the most severe losses, with QuIP\# reaching 12.3--15.3\% degradation on CodeLlama. For CodeLlama, the High bucket consistently shows higher degradation on McEval (\eg AWQ: 6.7\% vs.\ 0.0\%; AQLM: 16.7\% vs.\ 0.0\%), while for Qwen this pattern partially reverses---several techniques degrade \emph{more} on Low-entropy prompts (\eg AWQ: 16.7\% Low vs.\ 6.7\% High).

The correlation analysis quantifies this model-level divergence. For CodeLlama, five of six techniques show significant positive correlations: QuIP\# has the strongest signal ($r = +0.325$, $p < 0.001$ for entropy; $r = +0.321$, $p < 0.001$ for length), followed by BitsAndBytes ($r = +0.203$) and AWQ ($r = +0.192$). GPTQ is the sole exception. For Qwen, only two of twelve correlations reach significance, and one runs in the opposite direction---GPTQ shows a negative correlation ($r = -0.156$, $p < 0.001$), meaning simpler prompts degrade more. All other Qwen techniques show near-zero correlations. One possible explanation is that Qwen, trained on a larger and more diverse corpus, develops more redundant internal representations that are more resilient to precision loss---whereas CodeLlama may rely on less redundant weight configurations for complex reasoning.
\rev{Since the High-entropy bucket is dominated by BigCodeBench (Table~\ref{tab:entropy-buckets}), we recompute the correlation \emph{within BigCodeBench alone}: the model-level divergence persists---five of six CodeLlama techniques retain significance while Qwen remains largely insensitive---confirming the finding is not a benchmark artifact. Full results are in our replication package~\cite{replication}.}

The BCB-Python quality metrics (Table~\ref{tab:model-performance-python}) add a complementary perspective. For CodeLlama, the techniques that most aggressively quantize---QuIP\# and AQLM---not only degrade correctness on complex tasks but also inflate Reliability (43 and 32 vs.\ 22 in FP) and Maintainability (886 and 883 vs.\ 819), despite producing shorter code. Conversely, GGUF and BitsAndBytes reduce both LoC and Maintainability relative to FP, suggesting that their correctness losses are not accompanied by proportional quality degradation. For Qwen, the quantized variants are more tightly clustered around the baseline, with BitsAndBytes showing the largest divergence (LoC: 17,909 vs.\ 17,281; Reliability: 30 vs.\ 21). These patterns indicate that on complex benchmarks, correctness degradation and quality degradation do not always co-occur---a technique may preserve quality while losing correctness, or vice versa.

% AFTER
% \smallskip
% \begin{tcolorbox}[summarybox, title=\textbf{Summary -- RQ$_3$}]
% Complexity-induced degradation is model-dependent: CodeLlama-7B is sensitive (five of six techniques significant, QuIP\# strongest at $r \approx +0.32$), while Qwen2.5-Coder-7B is largely unaffected.
% \end{tcolorbox}

\smallskip
\begin{tcolorbox}[summarybox, title=\textbf{Summary -- RQ$_3$}]
Quantization robustness to input complexity is strongly model-dependent. For CodeLlama-7B, prompt complexity significantly correlates with degradation for five of six techniques, with QuIP\# showing the strongest sensitivity ($r \approx +0.32$, $p < 0.001$). For Qwen2.5-Coder-7B, almost no significant correlations exist---the model is largely unaffected by input complexity when quantized.
\end{tcolorbox}

\subsection{Qualitative Example}
\label{sec:qualitative-example}

\begin{figure*}[!t]
\centering
\includegraphics[width=\linewidth]{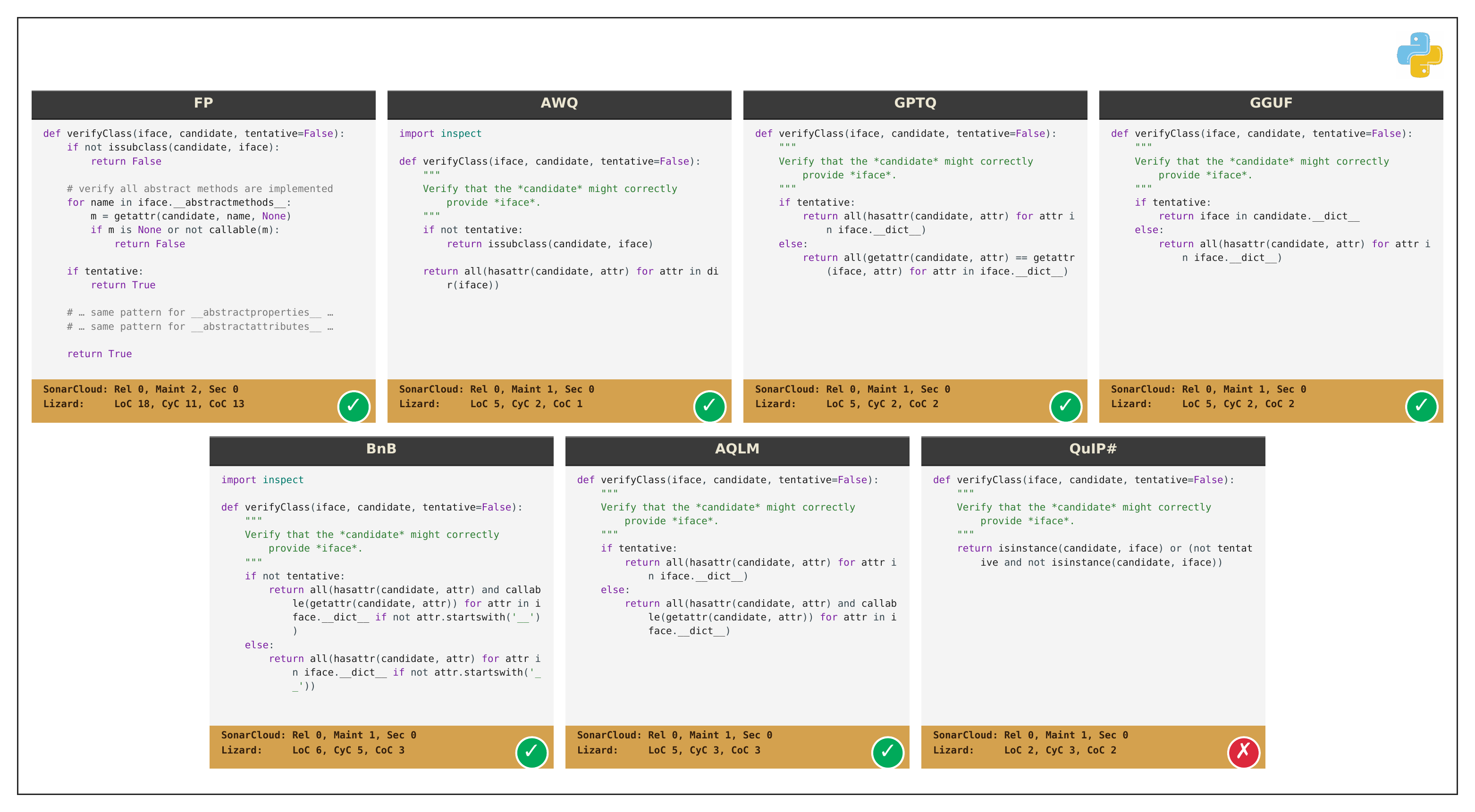}
\caption{Qualitative comparison of Qwen2.5-Coder-7B outputs across all seven configurations (FP + six quantized variants) for a single task. Each panel reports SonarCloud (Reliability, Maintainability, Security) and Lizard (LoC, CyC, CoC) metrics together with the pass ($\checkmark$) / fail ($\times$) outcome.}
\label{fig:qualitative-example}
\end{figure*}

\rev{To complement the aggregate statistics, Figure~\ref{fig:qualitative-example} shows Qwen2.5-Coder-7B's outputs for a single Python task---verifying that a candidate class implements an interface---under all seven configurations (FP + six quantized variants), annotated with the corresponding pass/fail outcome and SonarCloud/Lizard metrics. The example illustrates two patterns observed throughout our results. First, technique-level variation is visible at the source level: the FP baseline produces a structured 18-line routine that explicitly checks whether the candidate implements the interface's required abstract methods, while the abbreviated comments indicate the same checking pattern for abstract properties and attributes. In contrast, QuIP\# collapses the logic into a two-line expression (\texttt{isinstance(candidate, iface) or (not tentative and not isinstance(candidate, iface))}) that bypasses the abstract-method verification logic and is the only variant to fail the test suite---consistent with the QuIP\# degradation observed in RQ$_1$. Second, the remaining quantized variants (AWQ, GPTQ, GGUF, BitsAndBytes, AQLM) produce concise yet correct alternatives, reducing LoC from 18 to 5--6 and Cyclomatic Complexity from 11 to 2--5 while preserving functionality---consistent with the technique-dependent trade-offs reported in RQ$_2$.}

% \begin{figure}[!t]
% %\begin{figure*}[h]
% \centering
% \includegraphics[width=\linewidth]{sections/images/E1_final-2.pdf}
% \caption{Qualitative comparison of Qwen2.5-Coder-7B outputs across all seven configurations (FP + six quantized variants) for a single task. Each panel reports SonarCloud (Reliability, Maintainability, Security) and Lizard (LoC, CyC, CoC) metrics together with the pass ($\checkmark$) / fail ($\times$) outcome.}
% \label{fig:qualitative-example}
% \end{figure}

\section{\rev{Implications of our Findings}}
\label{sec:implications}

\subsection{\rev{Implications for Practitioners}}
\rev{
Our findings translate into concrete deployment guidance. Across both model families, 4-bit quantization yields a 63--66\% VRAM reduction (14.23\,GB $\rightarrow$ $\sim$5\,GB on Qwen2.5-Coder-7B), making local inference feasible on consumer GPUs without sacrificing aggregate correctness. Technique selection should follow the dominant deployment constraint:
}

\begin{itemize}
\item \rev{\textbf{Correctness-first:} AQLM is the only technique with a significant pass@1 \emph{improvement} (BCB-Python, OR\,=\,0.62), at the cost of higher latency.}
\item \rev{\textbf{Throughput-first:} GPTQ leads in throughput (1.65$\times$ FP16) while preserving correctness.}
\item \rev{\textbf{CPU-only:} GGUF remains the practical choice via the \texttt{llama.cpp} ecosystem.}
\item \rev{\textbf{Footprint-first:} QuIP\# achieves the smallest footprint but produces the only significant correctness degradations and the strongest complexity sensitivity ($r\,\approx\,+0.32$, CodeLlama). Use only when memory is binding and prompts are short.}
\item \rev{\textbf{Quality-sensitive:} AWQ inflates Java code smells (medium effect) and BitsAndBytes produces the largest Python complexity effect (CoC, large)---weigh these against their otherwise strong correctness profiles.}
\end{itemize}

\rev{\noindent Security attributes remain stable across all configurations.}

\subsection{\rev{Implications for Researchers}}

\rev{Three findings have broader implications. \textbf{First}, the technique-specific variation we report shows that single-technique evaluations cannot surface the trade-offs that emerge in comparative designs. \textbf{Second}, the model-dependence of complexity sensitivity (CodeLlama: 5/6 techniques significant; Qwen: largely insensitive) suggests representational redundancy from larger, more diverse pretraining corpora may act as implicit robustness to weight compression---a hypothesis warranting investigation. \textbf{Third}, RQ$_3$ introduces a previously uninvestigated evaluation axis: pass@1 averaged over a benchmark conceals systematic, complexity-correlated degradation that emerges only under stratified analysis.
}

\section{Threats to Validity}
\label{sec:threats}

\textbf{Construct validity.} We rely on SonarCloud as the sole static analysis tool, justified by prior evidence of strong consistency with language-specific alternatives~\cite{afrin2025quantization}. Different tools may capture additional quality dimensions. Our complexity characterization (token length, Shannon entropy) represents only two facets of prompt difficulty; other dimensions such as algorithmic complexity may also influence quantization sensitivity.

\textbf{Internal validity.} We chose two model families (Qwen2.5-Coder-7B, CodeLlama-7B) extensively used in prior quantization research~\cite{afrin2025quantization,giagnorio2025quantizing}, though results may differ for other architectures or scales. For quantization, we used pre-quantized Hugging Face checkpoints where available and quantized locally with WikiText-2 calibration for AQLM and QuIP\#---a standard choice~\cite{frantar2022gptq, lin2024awq}; code-specific calibration corpora could yield different results. Temperature was set to 0 for deterministic outputs; results under stochastic sampling may differ. 
\rev{We partially mitigated output variability via a ten-run check on Qwen2.5-Coder-7B-Instruct/McEval-Python; Friedman's test confirmed no significant run-level differences.}
\rev{For RQ$_3$, the High-entropy bucket is dominated by BigCodeBench; we mitigate this confound through a within-BigCodeBench replication (Section~\ref{sec:results}) that preserves the model-level divergence.}

\textbf{Conclusion validity.} We employ McNemar's test (binary correctness) and Wilcoxon signed-rank test (continuous quality metrics) with Holm-Bonferroni correction for multiple comparisons, supplemented by Cliff's delta effect sizes to distinguish statistical from practical significance.

\textbf{External validity.} Our findings are scoped to 7B models, two languages (Python, Java), three benchmarks, and 4-bit precision. Results may differ at other model scales, bit-widths, languages, or task types. RQ$_3$ is restricted to Python due to BigCodeBench's language coverage.

\section{Conclusion and Future Work}
\label{sec:conclusion}
We compared six weight-only quantization techniques---GPTQ, AWQ, QuIP\#, AQLM, BitsAndBytes, and GGUF---at 4-bit precision on Qwen2.5-Coder-7B and CodeLlama-7B, evaluating functional correctness, code quality, and robustness to input complexity. Our results show that 4-bit quantization largely preserves pass@1, but the choice of technique matters: AQLM consistently matches or exceeds the full-precision baseline, while QuIP\# produces the only significant correctness losses. Security is unaffected across all configurations, but AWQ significantly increases maintainability issues in Java (medium effect), and BitsAndBytes introduces the largest quality degradation on Python complexity metrics (large effect). Quantization robustness to prompt complexity is model-dependent: CodeLlama shows significant complexity--degradation correlations for five of six techniques ($r$ up to $+0.32$, $p<0.001$), while Qwen is largely insensitive.
\rev{Unlike prior work~\cite{afrin2025quantization} that assessed AWQ alone, our six-technique sweep reveals technique-specific trade-offs that single-technique studies cannot surface; the practical and research consequences are discussed in Section~\ref{sec:implications}.}

\textbf{Future work.} Several directions follow naturally from our results. First, extending the analysis to other precision levels (2--3\,bit) and larger model scales would test whether the technique-specific trade-offs reported here persist or invert. Second, the use of code-specific calibration datasets~\cite{giagnorio2025quantizing} may further mitigate the qualitative degradations we observe---notably AWQ on Java maintainability. Third, the interaction between weight quantization and inference-time optimizations such as speculative decoding, KV-cache quantization, and structured pruning remains an open question. Finally, the model-dependence of complexity sensitivity invites a controlled study isolating the role of pretraining-corpus diversity in shaping quantization robustness. A replication package containing all generated code, evaluation scripts, and detailed efficiency measurements is available at~\cite{replication}.

\section*{Acknowledgments}
The authors acknowledge the support of the National Science Foundation, which funded this research under grant NSF CCF-2451058.

%\balance
\bibliographystyle{IEEEtran}
\bibliography{main}

@article{crupi2025effectiveness,
  title={On the Effectiveness of LLM-as-a-judge for Code Generation and Summarization},
  author={Crupi, Giuseppe and Tufano, Rosalia and Velasco, Alejandro and Mastropaolo, Antonio and Poshyvanyk, Denys and Bavota, Gabriele},
  journal={IEEE Transactions on Software Engineering},
  year={2025},
  publisher={IEEE}
}

@inproceedings{lee2024owq,
  title={{OWQ}: Outlier-Aware Weight Quantization for Efficient Fine-Tuning and Inference of Large Language Models},
  author={Lee, Changhun and Jin, Jungyu and Kim, Taesu and Kim, Hyungjun and Park, Eunhyeok},
  booktitle={Proceedings of the AAAI Conference on Artificial Intelligence},
  volume={38},
  number={12},
  pages={13355--13364},
  year={2024}
}

@inproceedings{kim2024squeezellm,
  title={{SqueezeLLM}: Dense-and-Sparse Quantization},
  author={Kim, Sehoon and Hooper, Coleman and Gholami, Amir and Dong, Zhen and Li, Xiuyu and Shen, Sheng and Mahoney, Michael W. and Keutzer, Kurt},
  booktitle={Proceedings of the 41st International Conference on Machine Learning (ICML)},
  year={2024}
}

@inproceedings{chee2024quip,
  title={{QuIP}: 2-Bit Quantization of Large Language Models With Guarantees},
  author={Chee, Jerry and Cai, Yaohui and Kuleshov, Volodymyr and De Sa, Christopher M.},
  booktitle={Advances in Neural Information Processing Systems},
  volume={36},
  year={2024}
}

@misc{badri2023hqq,
  title={Half-Quadratic Quantization of Large Machine Learning Models},
  author={Badri, Hicham and Shaji, Appu},
  url={https://mobiusml.github.io/hqq_blog/},
  month={November},
  year={2023}
}

@inproceedings{yao2022zeroquant,
  title={{ZeroQuant}: Efficient and Affordable Post-Training Quantization for Large-Scale Transformers},
  author={Yao, Zhewei and Aminabadi, Reza Yazdani and Zhang, Minjia and Wu, Xiaoxia and Li, Conglong and He, Yuxiong},
  booktitle={Advances in Neural Information Processing Systems},
  volume={35},
  pages={27168--27183},
  year={2022}
}

@inproceedings{ashkboos2024quarot,
  title={{QuaRot}: Outlier-Free 4-Bit Inference in Rotated {LLMs}},
  author={Ashkboos, Saleh and Mohtashami, Amirkeivan and Croci, Maximilian L. and Li, Bo and Cameron, Pashmina and Jaggi, Martin and Alistarh, Dan and Hoefler, Torsten and Hensman, James},
  booktitle={Advances in Neural Information Processing Systems},
  volume={37},
  pages={100213--100240},
  year={2024}
}

@article{sun2024flatquant,
  title={{FlatQuant}: Flatness Matters for {LLM} Quantization},
  author={Sun, Yuxuan and Liu, Ruikang and Bai, Haoli and Bao, Han and Zhao, Kang and Li, Yang and Hu, Jiaqi and Yu, Xianzhi and Hou, Lu and Yuan, Chun and others},
  journal={arXiv preprint arXiv:2410.09426},
  year={2024}
}

@inproceedings{liu2024vptq,
  title={{VPTQ}: Extreme Low-Bit Vector Post-Training Quantization for Large Language Models},
  author={Liu, Yifei and Wen, Jicheng and Wang, Yang and Ye, Shengyu and Zhang, Li Lyna and Cao, Ting and Li, Cheng and Yang, Mao},
  booktitle={Proceedings of the 2024 Conference on Empirical Methods in Natural Language Processing (EMNLP)},
  year={2024}
}

@inproceedings{shao2024omniquant,
  title={{OmniQuant}: Omnidirectionally Calibrated Quantization for Large Language Models},
  author={Shao, Wenqi and Chen, Mengzhao and Zhang, Zhaoyang and Xu, Peng and Zhao, Lirui and Li, Zhiqian and Zhang, Kaipeng and Gao, Peng and Qiao, Yu and Luo, Ping},
  booktitle={International Conference on Learning Representations (ICLR)},
  year={2024}
}

@article{yuan2023rptq,
  title={{RPTQ}: Reorder-Based Post-Training Quantization for Large Language Models},
  author={Yuan, Zhihang and Niu, Liu and Liu, Jiawei and Liu, Wenyu and Wang, Xinggang and Shang, Yuzhang and Sun, Guangyu and Wu, Qiang and Wu, Jiaxiang and Wu, Bingzhe},
  journal={arXiv preprint arXiv:2304.01089},
  year={2023}
}

@inproceedings{shang2024pbllm,
  title={{PB-LLM}: Partially Binarized Large Language Models},
  author={Shang, Yuzhang and Yuan, Zhihang and Wu, Qiang and Dong, Zhen},
  booktitle={International Conference on Learning Representations (ICLR)},
  year={2024}
}

@article{frantar2022gptq,
  title={{GPTQ}: Accurate Post-Training Quantization for Generative Pre-Trained Transformers},
  author={Frantar, Elias and Ashkboos, Saleh and Hoefler, Torsten and Alistarh, Dan},
  journal={arXiv preprint arXiv:2210.17323},
  year={2022}
}

@article{lin2024awq,
  title={{AWQ}: Activation-Aware Weight Quantization for On-Device {LLM} Compression and Acceleration},
  author={Lin, Ji and Tang, Jiaming and Tang, Haotian and Yang, Shang and Chen, Wei-Ming and Wang, Wei-Chen and Xiao, Guangxuan and Dang, Xingyu and Gan, Chuang and Han, Song},
  journal={Proceedings of Machine Learning and Systems},
  volume={6},
  pages={87--100},
  year={2024}
}

@inproceedings{dettmers2023qlora,
  title={{QLoRA}: Efficient Finetuning of Quantized {LLMs}},
  author={Dettmers, Tim and Pagnoni, Artidoro and Holtzman, Ari and Zettlemoyer, Luke},
  booktitle={Advances in Neural Information Processing Systems},
  volume={36},
  pages={10088--10115},
  year={2023}
}

@inproceedings{dettmers2022gpt3,
  title={{LLM.int8()}: 8-bit Matrix Multiplication for Transformers at Scale},
  author={Dettmers, Tim and Lewis, Mike and Belkada, Younes and Zettlemoyer, Luke},
  booktitle={Advances in Neural Information Processing Systems},
  volume={35},
  pages={30318--30332},
  year={2022}
}

@inproceedings{egiazarian2024extreme,
  title={Extreme Compression of Large Language Models via Additive Quantization},
  author={Egiazarian, Vage and Panferov, Andrei and Kuznedelev, Denis and Frantar, Elias and Babenko, Artem and Alistarh, Dan},
  booktitle={Proceedings of the 41st International Conference on Machine Learning (ICML)},
  year={2024}
}

@inproceedings{tseng2024quip,
  title={{QuIP\#}: Even Better {LLM} Quantization with Hadamard Incoherence and Lattice Codebooks},
  author={Tseng, Albert and Chee, Jerry and Sun, Qingyao and Kuleshov, Volodymyr and De Sa, Christopher},
  booktitle={Proceedings of the 41st International Conference on Machine Learning (ICML)},
  volume={235},
  pages={48630--48656},
  year={2024},
  publisher={PMLR}
}

@inproceedings{dettmers2024spqr,
  title={{SpQR}: A Sparse-Quantized Representation for Near-Lossless {LLM} Weight Compression},
  author={Dettmers, Tim and Svirschevski, Ruslan and Egiazarian, Vage and Kuznedelev, Denis and Frantar, Elias and Ashkboos, Saleh and Borzunov, Alexander and Hoefler, Torsten and Alistarh, Dan},
  booktitle={International Conference on Learning Representations (ICLR)},
  year={2024}
}

@inproceedings{xiao2023smoothquant,
  title={{SmoothQuant}: Accurate and Efficient Post-Training Quantization for Large Language Models},
  author={Xiao, Guangxuan and Lin, Ji and Seznec, Mickael and Wu, Hao and Demouth, Julien and Han, Song},
  booktitle={International Conference on Machine Learning (ICML)},
  pages={38087--38099},
  year={2023},
  publisher={PMLR}
}

@article{liu2024llm,
  title={{LLM-QAT}: Data-Free Quantization Aware Training for Large Language Models},
  author={Liu, Zechun and Oguz, Barlas and Zhao, Changsheng and Chang, Ernie and Stock, Pierre and Mehdad, Yashar and Shi, Yangyang and Krishnamoorthi, Raghuraman and Chandra, Vikas},
  journal={arXiv preprint arXiv:2305.17888},
  year={2023}
}

@article{shen2024edgeqat,
  title={{EdgeQAT}: Entropy and Distribution Guided Quantization-Aware Training for the Acceleration of Lightweight {LLMs} on the Edge},
  author={Shen, Xuan and Zhao, Peiyan and Chen, Gong and Wang, Zhenglun and Lin, Yanzhi and others},
  journal={arXiv preprint arXiv:2402.10787},
  year={2024}
}

@inproceedings{shi2024greening,
  title={Greening Large Language Models of Code},
  author={Shi, Jieke and Yang, Zhou and Kang, Hong Jin and Xu, Bowen 
          and He, Junda and Lo, David},
  booktitle={Proceedings of the 46th International Conference on Software 
             Engineering: Software Engineering in Society (ICSE-SEIS)},
  pages={129--140},
  year={2024},
  organization={ACM},
  doi={10.1145/3639475.3640097}
}

@article{wang2021codet5,
  title={{CodeT5}: Identifier-Aware Unified Pre-Trained Encoder-Decoder Models for Code Understanding and Generation},
  author={Wang, Yue and Wang, Weishi and Joty, Shafiq and Hoi, Steven C. H.},
  journal={Proceedings of the 2021 Conference on Empirical Methods in Natural Language Processing},
  pages={8696--8708},
  year={2021}
}

@inproceedings{nijkamp2023codegen,
  title={{CodeGen}: An Open Large Language Model for Code with Multi-Turn Program Synthesis},
  author={Nijkamp, Erik and Pang, Bo and Hayashi, Hiroaki and Tu, Lifu and Wang, Huan and Zhou, Yingbo and Savarese, Silvio and Xiong, Caiming},
  booktitle={The Eleventh International Conference on Learning Representations (ICLR)},
  year={2023}
}

@article{li2022competition,
  title={Competition-Level Code Generation with {AlphaCode}},
  author={Li, Yujia and Choi, David and Chung, Junyoung and Kushman, Nate and Schrittwieser, Julian and Leblond, R{\'e}mi and Eccles, Tom and Keeling, James and Gimeno, Felix and Dal Lago, Agustin and others},
  journal={Science},
  volume={378},
  number={6624},
  pages={1092--1097},
  year={2022}
}

@article{fried2023incoder,
  title={{InCoder}: A Generative Model for Code Infilling and Synthesis},
  author={Fried, Daniel and Aghajanyan, Armen and Lin, Jessy and Wang, Sida and Wallace, Eric and Shi, Freda and Zhong, Ruiqi and Yih, Wen-tau and Zettlemoyer, Luke and Lewis, Mike},
  journal={The Eleventh International Conference on Learning Representations (ICLR)},
  year={2023}
}

@article{roziere2023code,
  title={Code {Llama}: Open Foundation Models for Code},
  author={Rozi{\`e}re, Baptiste and Gehring, Jonas and Gloeckle, Fabian and Sootla, Sten and Gat, Itai and Tan, Xiaoqing Ellen and Adi, Yossi and Liu, Jingyu and Remez, Tal and Rapin, J{\'e}r{\'e}my and others},
  journal={arXiv preprint arXiv:2308.12950},
  year={2023}
}

@article{touvron2023llama,
  title={{Llama 2}: Open Foundation and Fine-Tuned Chat Models},
  author={Touvron, Hugo and Martin, Louis and Stone, Kevin and Albert, Peter and Almahairi, Amjad and Babaei, Yasmine and Bashlykov, Nikolay and Batra, Soumya and Bhargava, Prajjwal and Bhosale, Shruti and others},
  journal={arXiv preprint arXiv:2307.09288},
  year={2023}
}

@article{hui2024qwen2,
  title={{Qwen2.5-Coder} Technical Report},
  author={Hui, Binyuan and Yang, Jian and Cui, Zeyu and Yang, Jiaxi and Liu, Dayiheng and Zhang, Lei and Liu, Tianyu and Zhang, Jiajun and Yu, Bowen and Dang, Kai and others},
  journal={arXiv preprint arXiv:2409.12186},
  year={2024}
}

@article{fan2023large,
  title={Large Language Models for Software Engineering: Survey and Open Problems},
  author={Fan, Angela and Gokkaya, Beliz and Harman, Mark and Lyubarskiy, Mitya and Sengupta, Shubho and Yoo, Shin and Zhang, Jie M.},
  journal={arXiv preprint arXiv:2310.03533},
  year={2023}
}

@article{tufano2020unit,
  title={Unit Test Case Generation with Transformers and Focal Context},
  author={Tufano, Michele and Watson, Cody and Bavota, Gabriele and Penta, Massimiliano Di and White, Martin and Poshyvanyk, Denys},
  journal={arXiv preprint arXiv:2009.05617},
  year={2020}
}

@inproceedings{ahmed2024automatic,
  title={Automatic semantic augmentation of language model prompts (for code summarization)},
  author={Ahmed, Toufique and Pai, Kunal Suresh and Devanbu, Premkumar and Barr, Earl},
  booktitle={Proceedings of the IEEE/ACM 46th international conference on software engineering},
  pages={1--13},
  year={2024}
}

@inproceedings{ahmed2022fewshot,
  title={Few-Shot Training {LLMs} for Project-Specific Code-Summarization},
  author={Ahmed, Toufique and Devanbu, Premkumar},
  booktitle={Proceedings of the 37th IEEE/ACM International Conference on Automated Software Engineering},
  pages={1--13},
  year={2022}
}

@article{zhuo2024bigcodebench,
  title={{BigCodeBench}: Benchmarking Code Generation with Diverse Function Calls and Complex Instructions},
  author={Zhuo, Terry Yue and Vu, Minh Chien and Chim, Jenny and Hu, Han and Yu, Wenhao and Widyasari, Ratnadira and Yusuf, Imam Nur Bani and Zhan, Haolan and He, Junda and Paul, Indraneil and others},
  journal={arXiv preprint arXiv:2406.15877},
  year={2024}
}

@article{quan2025codeelo,
  title={{CodeElo}: Benchmarking Competition-Level Code Generation of {LLMs} with Human-Comparable {Elo} Ratings},
  author={Quan, Shanghaoran and Ding, Jiaxi and Liu, Yibo and Tang, Zhiqiang and Ye, Han},
  journal={arXiv preprint arXiv:2501.01257},
  year={2025}
}

@inproceedings{castano2023exploring,
  title={Exploring the Carbon Footprint of {Hugging Face}'s {ML} Models: A Repository Mining Study},
  author={Casta{\~n}o, Josu{\'e} and Mart{\'i}nez-Fern{\'a}ndez, Silverio and Franch, Xavier and Bogner, Justus},
  booktitle={2023 ACM/IEEE International Symposium on Empirical Software Engineering and Measurement (ESEM)},
  pages={1--12},
  year={2023},
  organization={IEEE}
}

@article{patterson2021carbon,
  title={Carbon Emissions and Large Neural Network Training},
  author={Patterson, David and Gonzalez, Joseph and Le, Quoc and Liang, Chen and Munguia, Lluis-Miquel and Rothchild, Daniel and So, David and Texier, Maud and Dean, Jeff},
  journal={arXiv preprint arXiv:2104.10350},
  year={2021}
}

@article{novelli2024generative,
  title={Generative {AI} in {EU} Law: Liability, Privacy, Intellectual Property, and Cybersecurity},
  author={Novelli, Claudio and Casolari, Federico and Rotolo, Antonino and Taddeo, Mariarosaria and Floridi, Luciano},
  journal={arXiv preprint arXiv:2401.07348},
  year={2024}
}

@article{wu2024unveiling,
  title={Unveiling Security, Privacy, and Ethical Concerns of {ChatGPT}},
  author={Wu, Jiawen and Ouyang, Xiangnian and Chen, Haoyu and others},
  journal={Journal of Information and Intelligence},
  year={2024}
}

@inproceedings{houlsby2019parameter,
  title={Parameter-Efficient Transfer Learning for {NLP}},
  author={Houlsby, Neil and Giurgiu, Andrei and Jastrzebski, Stanislaw and Morrone, Bruna and De Laroussilhe, Quentin and Gesmundo, Andrea and Attariyan, Mona and Gelly, Sylvain},
  booktitle={International Conference on Machine Learning (ICML)},
  pages={2790--2799},
  year={2019}
}

@article{hu2022lora,
  title={{LoRA}: Low-Rank Adaptation of Large Language Models},
  author={Hu, Edward J. and Shen, Yelong and Wallis, Phillip and Allen-Zhu, Zeyuan and Li, Yuanzhi and Wang, Shean and Wang, Lu and Chen, Weizhu},
  journal={The Tenth International Conference on Learning Representations (ICLR)},
  year={2022}
}

@article{lester2021power,
  title={The Power of Scale for Parameter-Efficient Prompt Tuning},
  author={Lester, Brian and Al-Rfou, Rami and Constant, Noah},
  journal={Proceedings of the 2021 Conference on Empirical Methods in Natural Language Processing},
  pages={3045--3059},
  year={2021}
}

@article{li2021prefix,
  title={Prefix-Tuning: Optimizing Continuous Prompts for Generation},
  author={Li, Xiang Lisa and Liang, Percy},
  journal={Proceedings of the 59th Annual Meeting of the Association for Computational Linguistics},
  pages={4582--4597},
  year={2021}
}

@article{wang2023one,
  title={One Adapter for All Programming Languages? Adapter Tuning for Code Search and Summarization},
  author={Wang, Deze and Tan, Zhouyang and Chen, Renyu and Zhang, Junjie},
  journal={Proceedings of the IEEE/ACM 45th International Conference on Software Engineering},
  year={2023}
}

@inproceedings{weyssow2023exploring,
  title={Exploring Parameter-Efficient Fine-Tuning Techniques for Code Generation with Large Language Models},
  author={Weyssow, Martin and Zhou, Xin and Kim, Kisub and Lo, David and Sahraoui, Houari},
  booktitle={arXiv preprint arXiv:2308.10462},
  year={2023}
}

@inproceedings{liu2023empirical,
  title={An Empirical Study of Parameter-Efficient Fine-Tuning Methods for Pre-Trained Code Models},
  author={Liu, Jiaxing and Keung, Jacky and Zhou, Qiang and Liao, Yue},
  booktitle={2023 38th IEEE/ACM International Conference on Automated Software Engineering (ASE)},
  pages={397--408},
  year={2023}
}

@inproceedings{ayupov2022parameter,
  title={Parameter-Efficient Fine-Tuning for Pre-Trained Code Models},
  author={Ayupov, Shamil and Ren, Shuo},
  booktitle={Proceedings of the 1st International Workshop on Natural Language-based Software Engineering},
  year={2022}
}

@article{hsieh2023distilling,
  title={Distilling Step-by-Step! Outperforming Larger Language Models with Less Training Data and Smaller Model Sizes},
  author={Hsieh, Cheng-Yu and Li, Chun-Liang and Yeh, Chih-Kuan and Nakhost, Hootan and Fujii, Yasuhisa and Ratner, Alex and Krishna, Ranjay and Lee, Chen-Yu and Pfister, Tomas},
  journal={Findings of the Association for Computational Linguistics: ACL 2023},
  pages={8003--8017},
  year={2023}
}

@article{chaudhary2023code,
  title={Code {Alpaca}: An Instruction-Following {LLaMA} Model for Code Generation},
  author={Chaudhary, Sahil},
  journal={GitHub repository},
  year={2023}
}

@article{wei2023magicoder,
  title={{Magicoder}: Source Code Is All You Need},
  author={Wei, Yuxiang and Wang, Zhe and Liu, Jiawei and Ding, Yifeng and Zhang, Lingming},
  journal={arXiv preprint arXiv:2312.02120},
  year={2023}
}

@article{daloisio2024compression,
  title={On the Compression of Natural Language Models},
  author={D'Aloisio, Giordano and Di Marco, Antinisca and Di Stasi, Andrea and Ferrara, Andrea},
  journal={arXiv preprint arXiv:2404.09095},
  year={2024}
}

@incollection{gholami2022survey,
  title={A Survey of Quantization Methods for Efficient Neural Network Inference},
  author={Gholami, Amir and Kim, Sehoon and Dong, Zhen and Yao, Zhewei and Mahoney, Michael W. and Keutzer, Kurt},
  booktitle={Low-Power Computer Vision},
  pages={291--326},
  year={2022},
  publisher={Chapman and Hall/CRC}
}

@article{wang2024survey,
  title={A Survey on Model Compression for Large Language Models},
  author={Wang, Xunyu and Li, Jian and Liu, Yong and Ma, Can and Wang, Weiping},
  journal={arXiv preprint arXiv:2308.07633},
  year={2024}
}

@article{esser2020learned,
  title={Learned Step Size Quantization},
  author={Esser, Steven K. and McKinstry, Jeffrey L. and Bablani, Deepika and Appuswamy, Rathinakumar and Modha, Dharmendra S.},
  journal={arXiv preprint arXiv:1902.08153},
  year={2020}
}

@inproceedings{cai2020zeroq,
  title={{ZeroQ}: A Novel Zero Shot Quantization Framework},
  author={Cai, Yaohui and Yao, Zhewei and Dong, Zhen and Gholami, Amir and Mahoney, Michael W. and Keutzer, Kurt},
  booktitle={Proceedings of the IEEE/CVF Conference on Computer Vision and Pattern Recognition},
  pages={13169--13178},
  year={2020}
}

@book{cochran1954some,
  title={Some Methods for Strengthening the Common $\chi^2$ Tests},
  author={Cochran, William G.},
  journal={Biometrics},
  volume={10},
  number={4},
  pages={417--451},
  year={1954}
}

@article{mann1947test,
  title={On a Test of Whether One of Two Random Variables is Stochastically Larger than the Other},
  author={Mann, Henry B. and Whitney, Donald R.},
  journal={The Annals of Mathematical Statistics},
  volume={18},
  number={1},
  pages={50--60},
  year={1947}
}

@book{cohen1988statistical,
  title={Statistical Power Analysis for the Behavioral Sciences},
  author={Cohen, Jacob},
  edition={2nd},
  publisher={Lawrence Erlbaum Associates},
  year={1988}
}

@article{jin2024comprehensive,
  title={A Comprehensive Evaluation of Quantization Strategies for Large Language Models},
  author={Jin, Renren and Du, Jiangcun and Huang, Wuwei and Liu, Wei and Luan, Jian and Wang, Bin and Xiong, Deyi},
  journal={Proceedings of the 62nd Annual Meeting of the Association for Computational Linguistics (ACL)},
  year={2024}
}

@article{yetistiren2023evaluating,
  title={Evaluating the Code Quality of {AI}-Assisted Code Generation Tools: An Empirical Study on {GitHub Copilot}, {Amazon CodeWhisperer}, and {ChatGPT}},
  author={Yeti{\c{s}}tiren, Burak and {\"O}zsoy, I{\c{s}}{\i}l and Ayerdem, Miray and T{\"u}z{\"u}n, Eray},
  journal={arXiv preprint arXiv:2304.10778},
  year={2023}
}

@inproceedings{yu2024codereval,
  title={{CoderEval}: A Benchmark of Pragmatic Code Generation with Generative Pre-Trained Models},
  author={Yu, Hao and Shen, Bo and Ran, Dezhi and Zhang, Jiaxin and Zhang, Qi and Ma, Yuchi and Liang, Guangtai and Li, Ying and Wang, Qianxiang and Xie, Tao},
  booktitle={Proceedings of the IEEE/ACM 46th International Conference on Software Engineering (ICSE)},
  pages={1--13},
  year={2024},
  doi={10.1145/3597503.3623316}
}

@inproceedings{chai2024mceval,
  title={{McEval}: Massively Multilingual Code Evaluation},
  author={Chai, Linzheng and Liu, Shukai and Yang, Jian and Yin, Yuwei and Jin, Ke and Liu, Jiaheng and Sun, Tao and Zhang, Ge and Ren, Changyu and Guo, Hongcheng and others},
  booktitle={arXiv preprint arXiv:2406.07436},
  year={2024}
}

@article{li2023structured,
  title={Structured Chain-of-Thought Prompting for Code Generation},
  author={Li, Jia and Li, Ge and Li, Yongmin and Jin, Zhi},
  journal={ACM Transactions on Software Engineering and Methodology},
  year={2023}
}

@inproceedings{coignion2024performance,
  title={A Performance Study of {LLM}-Generated Code on {LeetCode}},
  author={Coignion, Tristan and Quinton, Cl{\'e}ment and Rouvoy, Romain},
  booktitle={Proceedings of the 28th International Conference on Evaluation and Assessment in Software Engineering},
  pages={79--89},
  year={2024}
}

@article{ren2024reflectioncoder,
  title={{ReflectionCoder}: Learning from Reflection Sequence for Enhanced One-Off Code Generation},
  author={Ren, Houxing and Zhan, Mingjie and Wu, Zhongyuan and Zhou, Aojun and Pan, Junting and Li, Hongsheng},
  journal={arXiv preprint arXiv:2405.17057},
  year={2024}
}

@article{afrin2025resource,
  title={Resource-Efficient \& Effective Code Summarization},
  author={Afrin, Saima and Call, Jordan and Nguyen, Khanh-Nam and Chaparro, Oscar and Mastropaolo, Antonio},
  journal={arXiv preprint arXiv:2502.03617},
  year={2025}
}

@article{li2023instructcoder,
  title={{InstructCoder}: Instruction Tuning Large Language Models for Code Editing},
  author={Li, Kaixin and Hu, Qisheng and Zhao, Xu and Chen, Hui and Xie, Yuxi and Liu, Tiedong and Xie, Qizhe and He, Junxian},
  journal={arXiv preprint arXiv:2310.20329},
  year={2023}
}

@misc{sonarcloud,
  title={{SonarCloud}},
  author={{SonarSource}},
  howpublished={\url{https://docs.sonarsource.com/sonarqube-cloud/}},
  year={2025},
  note={Accessed: 2025-03-03}
}

@article{chen:arxiv2021,
  title={Evaluating Large Language Models Trained on Code},
  author={Chen, Mark and Tworek, Jerry and Jun, Heewoo and Yuan, Qiming and Pinto, Henrique Ponde de Oliveira and Kaplan, Jared and Edwards, Harri and Burda, Yuri and Joseph, Nicholas and Brockman, Greg and others},
  journal={arXiv preprint arXiv:2107.03374},
  year={2021}
}

@article{merity2016pointer,
  title={Pointer Sentinel Mixture Models},
  author={Merity, Stephen and Xiong, Caiming and Bradbury, James and Socher, Richard},
  journal={arXiv preprint arXiv:1609.07843},
  year={2016}
}

@inproceedings{wang2023review,
  title={A Review on Code Generation with {LLMs}: Application and Evaluation},
  author={Wang, Jian and Chen, Yufei},
  booktitle={2023 IEEE International Conference on Medical Artificial Intelligence (MedAI)},
  pages={284--289},
  year={2023},
  organization={IEEE}
}

@article{fakhoury2024llm,
  title={{LLM}-Based Test-Driven Interactive Code Generation: User Study and Empirical Evaluation},
  author={Fakhoury, Sarah and Naik, Aaditya and Sakkas, Georgios and Chakraborty, Saikat and Lahiri, Shuvendu K.},
  journal={IEEE Transactions on Software Engineering},
  year={2024}
}

@article{munoz2020empirical,
  title={An Empirical Validation of Cognitive Complexity as a Measure of Source Code Understandability},
  author={Mu{\~n}oz Bar{\'o}n, Marvin and Wyrich, Marvin and Wagner, Stefan},
  journal={Proceedings of the 14th ACM/IEEE International Symposium on Empirical Software Engineering and Measurement (ESEM)},
  pages={1--12},
  year={2020}
}

@article{friedman1937use,
  title={The Use of Ranks to Avoid the Assumption of Normality Implicit in the Analysis of Variance},
  author={Friedman, Milton},
  journal={Journal of the American Statistical Association},
  volume={32},
  number={200},
  pages={675--701},
  year={1937}
}

@book{Cliff:2005,
  title={Effect Sizes for Research: A Broad Practical Approach},
  author={Grissom, Robert J. and Kim, John J.},
  edition={2nd},
  publisher={Lawrence Erlbaum Associates},
  year={2005}
}

@article{mcnemar1947note,
  title={Note on the Sampling Error of the Difference Between Correlated Proportions or Percentages},
  author={McNemar, Quinn},
  journal={Psychometrika},
  volume={12},
  number={2},
  pages={153--157},
  year={1947}
}

@article{wilcoxon1945individual,
  title={Individual Comparisons by Ranking Methods},
  author={Wilcoxon, Frank},
  journal={Biometrics Bulletin},
  volume={1},
  number={6},
  pages={80--83},
  year={1945}
}

@article{holm1979simple,
  title={A Simple Sequentially Rejective Multiple Test Procedure},
  author={Holm, Sture},
  journal={Scandinavian Journal of Statistics},
  volume={6},
  number={2},
  pages={65--70},
  year={1979}
}

@article{shannon1948mathematical,
  title={A mathematical theory of communication},
  author={Shannon, Claude E},
  journal={The Bell system technical journal},
  volume={27},
  number={3},
  pages={379--423},
  year={1948},
  publisher={Nokia Bell Labs}
}

@article{nyamsuren2025evaluating,
  title={Evaluating quantized large language models for code generation on low-resource language benchmarks},
  author={Nyamsuren, Enkhbold},
  journal={Journal of Computer Languages},
  pages={101351},
  year={2025},
  publisher={Elsevier}
}

@article{ahmad2021unified,
  title={Unified pre-training for program understanding and generation},
  author={Ahmad, Wasi Uddin and Chakraborty, Saikat and Ray, Baishakhi and Chang, Kai-Wei},
  journal={arXiv preprint arXiv:2103.06333},
  year={2021}
}

@article{flake8,
  title={Flake8: Your tool for style guide enforcement. 2021},
  author={Ziad{\'e}, T and Cordasco, I},
  journal={URL: http://flake8. pycqa. org (besucht am 27. 05. 2019)}
}

@article{giagnorio2025quantizing,
  title={Quantizing large language models for code generation: A differentiated replication},
  author={Giagnorio, Alessandro and Mastropaolo, Antonio and Afrin, Saima and Di Penta, Massimiliano and Bavota, Gabriele},
  journal={arXiv preprint arXiv:2503.07103},
  year={2025}
}

@inproceedings{codereval,
	title={Codereval: A benchmark of pragmatic code generation with generative pre-trained models},
	author={Yu, Hao and Shen, Bo and Ran, Dezhi and Zhang, Jiaxin and Zhang, Qi and Ma, Yuchi and Liang, Guangtai and Li, Ying and Wang, Qianxiang and Xie, Tao},
	booktitle={Proceedings of the 46th IEEE/ACM International Conference on Software Engineering},
	pages={1--12},
	year={2024}
}

@misc{ollama_repo,
  author       = {{Ollama Project}},
  title        = {Ollama},
  howpublished = {\url{https://github.com/ollama/ollama}},
  note         = {Accessed: 2026-05-08},
  year         = {2024}
}

@misc{thebloke_hf,
  author       = {TheBloke},
  title        = {TheBloke -- {H}ugging {F}ace model repository},
  howpublished = {\url{https://huggingface.co/TheBloke}},
  note         = {Accessed: 2026-05-08},
  year         = {2023}
}

@inproceedings{dettmers2023spqr,
  author       = {Dettmers, Tim and Svirschevski, Ruslan and Egiazarian, Vage and Kuznedelev, Denis and Frantar, Elias and Ashkboos, Saleh and Borzunov, Alexander and Hoefler, Torsten and Alistarh, Dan},
  title        = {{SpQR}: A Sparse-Quantized Representation for Near-Lossless {LLM} Weight Compression},
  booktitle    = {The Twelfth International Conference on Learning Representations (ICLR)},
  year         = {2024}
}

@misc{quip_for_all,
  author       = {Chu, Tianxiang},
  title        = {{QuIP}-for-all: Unified {QuIP\#} implementation supporting diverse architectures},
  howpublished = {\url{https://github.com/chu-tianxiang/QuIP-for-all}},
  note         = {Accessed: 2026-05-08},
  year         = {2024}
}

@misc{gguf_spec,
  author       = {Gerganov, Georgi},
  title        = {{GGUF} format specification},
  howpublished = {\url{https://github.com/ggerganov/ggml/blob/master/docs/gguf.md}},
  note         = {Accessed: 2026-05-08},
  year         = {2023}
}

@article{liu2024fail,
  title={Where Do Large Language Models Fail When Generating Code?},
  author={Liu, Zhijie and Tang, Yutian and Luo, Xiapu and Zhou, Yuming and Zhang, Liang Feng},
  journal={arXiv preprint arXiv:2406.08731},
  year={2024}
}

@inproceedings{genzel2002entropy,
  title={Entropy rate constancy in text},
  author={Genzel, Dmitriy and Charniak, Eugene},
  booktitle={Proceedings of the 40th Annual Meeting of the Association for Computational Linguistics (ACL)},
  pages={199--206},
  year={2002}
}

@article{siddiq2023generate,
  title={Generate and pray: Using sallms to evaluate the security of llm generated code},
  author={Siddiq, Mohammed Latif and Santos, Joanna CS},
  journal={arXiv preprint arXiv:2311.00889},
  year={2023}
}

@inproceedings{strubell2020energy,
  title={Energy and policy considerations for modern deep learning research},
  author={Strubell, Emma and Ganesh, Ananya and McCallum, Andrew},
  booktitle={Proceedings of the AAAI conference on artificial intelligence},
  volume={34},
  number={09},
  pages={13693--13696},
  year={2020}
}

@inproceedings{wei2023towards,
  title={Towards greener yet powerful code generation via quantization: An empirical study},
  author={Wei, Xiaokai and Gonugondla, Sujan Kumar and Wang, Shiqi and Ahmad, Wasi and Ray, Baishakhi and Qian, Haifeng and Li, Xiaopeng and Kumar, Varun and Wang, Zijian and Tian, Yuchen and others},
  booktitle={Proceedings of the 31st ACM Joint European Software Engineering Conference and Symposium on the Foundations of Software Engineering},
  pages={224--236},
  year={2023}
}

@article{kharma2025security,
  title={Security and Quality in LLM-Generated Code: A Multi-Language, Multi-Model Analysis},
  author={Kharma, Mohammed and Choi, Soohyeon and AlKhanafseh, Mohammed and Mohaisen, David},
  journal={arXiv preprint arXiv:2502.01853},
  year={2025}
}

@misc{PMD,
  title        = {PMD - Source Code Analyzer},
  author       = {PMD Development Team},
  year         = {2025},
  url          = {https://pmd.github.io},
  note         = {Static code analysis tool for Java and other languages}
}

@inproceedings{siddiq2024quality,
  title={Quality assessment of chatgpt generated code and their use by developers},
  author={Siddiq, Mohammed Latif and Roney, Lindsay and Zhang, Jiahao and Santos, Joanna Cecilia Da Silva},
  booktitle={Proceedings of the 21st International Conference on Mining Software Repositories},
  pages={152--156},
  year={2024}
}

@inproceedings{Mastropaolo:icse2023,
  author       = {Antonio Mastropaolo and
                  Luca Pascarella and
                  Emanuela Guglielmi and
                  Matteo Ciniselli and
                  Simone Scalabrino and
                  Rocco Oliveto and
                  Gabriele Bavota},
  title        = {On the Robustness of Code Generation Techniques: An Empirical Study
                  on GitHub Copilot},
  booktitle    = {45th {IEEE/ACM} International Conference on Software Engineering,
                  {ICSE} 2023, Melbourne, Australia, May 14-20, 2023},
  pages        = {2149--2160},
  publisher    = {{IEEE}},
  year         = {2023}
}

@misc{pylint,
    author = {PylintTeam},
    title  = {Pylint - code analysis for Python},
    howpublished = {\url{https://www.pylint.org/}},
    note         = {Accessed: 2025-03-03}
}

@misc{hinton2015distillingknowledge,
	title={Distilling the Knowledge in a Neural Network}, 
	author={Geoffrey Hinton and Oriol Vinyals and Jeff Dean},
	year={2015},
	eprint={1503.02531},
	archivePrefix={arXiv},
	primaryClass={stat.ML},
	url={https://arxiv.org/abs/1503.02531}, 
}

@misc{replication,
  title        = {Replication Package},
  howpublished = {\url{https://github.com/empirical-quant-project/empirical-quantization-study}},
  year         = {2026}
}

@inproceedings{afrin2025quantization,
  title={Is Quantization a Deal-breaker? Empirical Insights from Large Code Models},
  author={Afrin, Saima and Xu, Bowen and Mastropaolo, Antonio},
  booktitle={2025 IEEE International Conference on Software Maintenance and Evolution (ICSME)},
  pages={1--13},
  year={2025},
  organization={IEEE}
}

@article{shi2024efficient,
  title={Efficient and green large language models for software engineering: Vision and the road ahead},
  author={Shi, Jieke and Yang, Zhou and Lo, David},
  journal={ACM Transactions on Software Engineering and Methodology},
  year={2024},
  publisher={ACM New York, NY}
}

@article{liu2024refining,
  title={Refining chatgpt-generated code: Characterizing and mitigating code quality issues},
  author={Liu, Yue and Le-Cong, Thanh and Widyasari, Ratnadira and Tantithamthavorn, Chakkrit and Li, Li and Le, Xuan-Bach D and Lo, David},
  journal={ACM Transactions on Software Engineering and Methodology},
  volume={33},
  number={5},
  pages={1--26},
  year={2024},
  publisher={ACM New York, NY}
}

\end{document}